\documentclass[aps, rmp, twocolumn, showpacs,preprintnumbers,amsmath,amssymb]{revtex4}
\usepackage{textcomp}
\usepackage{amsmath,amssymb,amsfonts}
\usepackage{algorithmic}
\usepackage{graphicx}
\usepackage{epstopdf}
\usepackage{array}
\usepackage{dcolumn}
\usepackage{bm}
\usepackage{color}
\usepackage{longtable}

\newcolumntype{L}{>{\raggedright\arraybackslash}p}
\newcolumntype{C}{>{\centering\arraybackslash}m}

\begin{document}
\title{Engineering physics of superconducting hot-electron bolometer mixers}

\author{T.~M.~Klapwijk} 
\affiliation{Kavli Institute of Nanoscience, Faculty of Applied Sciences,
Delft University of Technology, Lorentzweg 1, 2628 CJ Delft, The Netherlands}
\affiliation{ Physics Department, Moscow State University of Education, 1 Malaya Pirogovskaya st., Moscow 119992, Russia}
\author{A.~V.~Semenov}
\affiliation{Physics Department, Moscow State University of Education, 1 Malaya Pirogovskaya st., Moscow 119992, Russia}
\affiliation{ Moscow Institute of Physics and Technology, Dolgoprudny, Russia}

\begin{abstract}
\boldmath
Superconducting hot-electron bolometers are presently the best performing mixing devices for the frequency range beyond {1.2 THz, where good quality superconductor-insulator-superconductor (SIS) devices do not exist.} Their physical appearance is very simple: an antenna consisting of a normal metal, sometimes a normal metal-superconductor bilayer, connected to a thin film of a narrow, short superconductor with a high resistivity in the normal state. The device is brought into an optimal operating regime by applying a dc current and a certain amount of local-oscillator power. Despite this technological simplicity its operation has found to be controlled by many different aspects of superconductivity,  all occurring simultaneously.  A core ingredient is the understanding that there are two sources of resistance in a superconductor: a charge conversion resistance occurring at an normal-metal-superconductor interface and a resistance due to time-dependent changes of the superconducting phase. The latter is responsible for the actual mixing process in a non-uniform superconducting environment set up by the bias-conditions and the geometry. The present understanding indicates that further improvement needs to be found in the use of other materials with a faster energy-relaxation rate.  Meanwhile several empirical parameters have become physically meaningful indicators of the devices, which will facilitate the technological developments. 
\end{abstract}

\maketitle
\tableofcontents{}

\section{\label{sec:level1}Introduction\protect}
\label{intro}

\begin{figure}[t]
\begin{center}
\includegraphics[width=0.7\columnwidth]{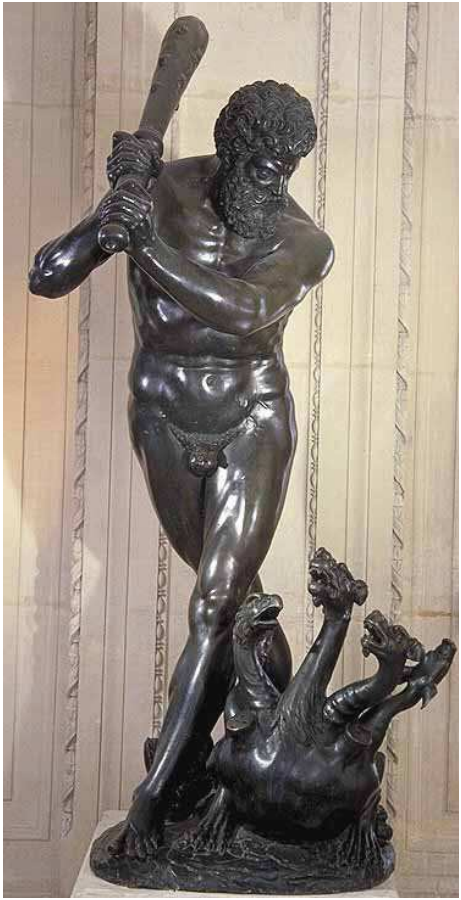}
\end{center}
\caption{\label{fig:hydra} Statue of Hercules fighting the snake Hydra. Each head that was hit led to the appearance of two new heads, symbolising the complexity of the physics of hot-electron bolometers. (Louvrecollection)}
\end{figure}
In the past 50 years astrochemistry has expanded enormously using high-resolution spectroscopy in the THz frequency range, provided by heterodyne instruments. A key role is played by low-noise mixing devices. A 1982 review on heterodyne mixing  for astronomy \cite{PhillipsWoody1982} marks the turning point, after the first decade, when semiconducting components are  being replaced by superconducting components. For the Atacama Large Millimeter Array (ALMA), based on superconductor-insulator-superconductor (SIS) tunnel junctions, the device-physics has emerged at the end of the 70-ies, early 80-ies and is summarized very effectively in a by now classic paper \cite{Tucker1985}. Since a good quality non-linear tunnelling device is needed the frequency range to which this technology can be extended is limited by materials, which enable good tunnel barriers. For practical superconducting devices niobium-based trilayer technology with proximitized aluminium covered with an aluminium-oxide barrier has been dominating since 1983 \cite{Gurvitch1983}.
{This niobium-work at Bell Laboratories, the parting shot of the prior lead-based digital Josephson-computer era, was ideally suited to take the pioneering SIS-work based on Pb tunnel-junctions to the practical use at astronomical telescopes including those in space \cite{Klapwijk2012}. The niobium tunnel-junctions} were equipped with low loss NbTiN or Al striplines, which served as antennas and impedance-matching structures. The maximum frequency range of this technology is about 1.2 THz, which was actually used in the Heterodyne Instrument for the Far-Infrared (HIFI) as used in the Herschel Space telescope \cite{Graauw2010}. In order to go beyond this frequency range the only suitable candidate known at the moment is the superconducting hot-electron bolometer (HEB). This device has been used, {mostly for astronomy - in HIFI up to 1.9 THz and to an impressive 4.7 THz on SOFIA \cite{Buechel2015} -} but also for near-field nanoscopy \cite{Keilmann2008}.

Compared to SIS the HEB-devices have developed much more slowly. An important practical constraint is that the higher frequencies can only be accessed from space. At the time when the Herschel Space telescope was conceived the HEB technology was still very immature and the extension up to 1.9 THz was already considered a bold step. Apart from space-based use there is no evident advantage of HEB mixers at lower frequencies, where SIS mixers are superior and hence dominate the instrumental development. A 2nd reason is that in comparison to SIS the physics of HEB devices is much more complex. In SIS tunnelling devices the system can be broken down into well-defined parts. Two bulk superconducting electrodes are weakly coupled by an insulating tunnel-barrier. The weakness of the coupling means that the insulating state and the superconducting state do not effect each others properties. Through the tunnel barrier a small current flows, which has the desired non-linear dependence on voltage, but is much smaller than a current that would influence the superconductor itself. In contrast, with superconducting HEB-devices (Fig.~\ref{fig:3Dview})  a number of differences occur. First of all, in view of the applications at frequencies higher than the superconducting energy gap the superconductor absorbs the radiation, in fact one uses the absorption of radiation by the superconductor to heat the electron-system in the superconductor. In order to do that effectively the superconductor is coupled to good conducting normal antennas, usually made of gold. Hence, the device is essentially a normal-metal-superconductor-normal-metal (NSN) device. As a consequence, although the SIS device embodied a weakly coupled system the HEB embodies a strongly coupled system, for which the properties are position-dependent. It means that the proximity-effect plays a role as well as the conversion of normal current to supercurrent, a subject which is part of the field of nonequilibrium superconductivity. In addition, the superconductor should be resistive, which involves vortex-physics. In developing the understanding of the physics of hot-electron bolometers, the simplicity of the device has turned into a multi-headed snake, which could only be conquered by the Herculean task of the superconducting community 
(Fig.~\ref{fig:hydra}).   

In recent years, a lot of progress has been made in extending the operating range of HEB's as well as in clarifying  the processes determining the physics of the HEB's. This review is intended to summarise the status of the field from both of these viewpoints. Such a review seems timely, because an important goal for the nearby future is the extension of heterodyne observations into the SuperTHz range. Several options are available ranging from stratospheric balloons in Antarctica, instruments on the airborne SOFIA-observatory and, most challenging, an observatory to be built at, for example, {Dome A in Antarctica} \cite{ShiDomeA2016}. Unfortunately, an analysis of the various noise sources could not be presented in this review, because systematic studies are not yet available. An exception is the subject of quantum noise, as treated in \cite{kollberg2006}, which has received experimental support in work of Zhang et al \cite{zhang2010}. The present review might stimulate further work in this direction.

\section{Brief description of standard devices}

\begin{figure}[t]
\begin{center}
\includegraphics[width=0.8\columnwidth]{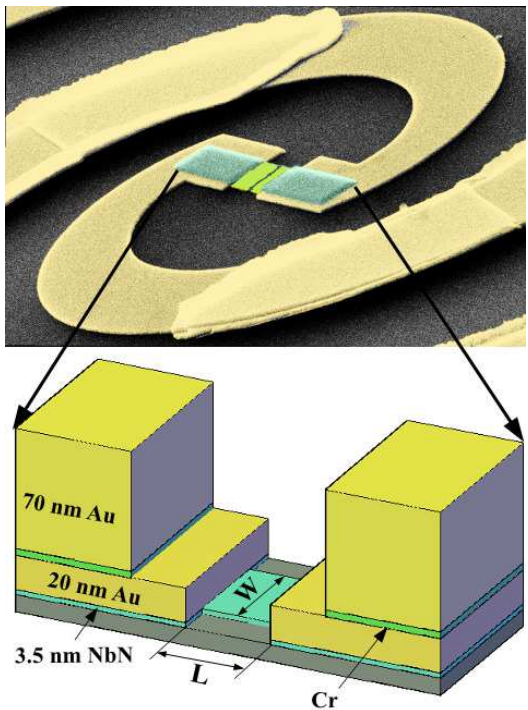}
\end{center}
\caption{\label{fig:3Dview} Upper panel: false color image of a hot-electron bolometer in the center of a spiral antenna of gold (Au). Lower panel:  3D view of a typical device. The green central part is a thin film of niobium-nitride and yellow indicates gold. Between the two gold films a thin adhesion layer of chromium (Cr) is inserted. {(Reproduced with permission from  \cite{Shcherbatenko2016})}} 
\end{figure}
A typical hot-electron bolometer mixer (Fig.~\ref{fig:3Dview}) consists of a centre piece of a thin film of a superconductor, usually niobium-nitride (NbN), although niobium-titanium-nitride (NbTiN) has been used as well \cite{MunozJacobs2006}. In practice \cite{Tretyakov2011}, the devices are made of a 3 to 4 nm thick film shaped to a width $W$ and length $L$. The aspect ratio is determined by the desired impedance taking the normal state resistance of the superconducting film as a guideline. As illustrated in the figure, the devices consist of 2 more crucial parts. In Fig.~\ref{fig:3Dview}, the central, bare NbN film, is connected to a NbN-Au bilayer, which on its turn is connected to a NbN-Au-Cr-Au multilayer, which serves as an antenna (a broadband spiral antenna). The sputter-deposition of the NbN is done while the oxidized silicon or quartz substrate is at a high temperature of 800 $^{\circ}$C. After cooling the substrate to 300 $^{\circ}$C a thin film of 20 nm thick Au is deposited \emph{in situ}.  This \emph{in situ} process is used to guarantee a good metallic contact between Au and NbN. The Au is subsequently locally removed by ion-etching and wet chemical etching through a window in an electron-beam resist, to define the actual active NbN part.  An additional layer of 70 nm of Au is deposited for the antenna. This 2nd Au layer is deposited in a separate deposition run, using a few nanometers of Cr for better adhesion. Obviously, the device is in principle a NSN device, with S the uncovered active part consisting of the very thin NbN film. The N-part is meant to provide a good dc electrical contact, which can be achieved in a variety of ways. It is shaped as an antenna to sense the radiation from free space, a waveguide  or a quasi-optical lens. Since the normal film N is in contact with a superconductor it may also be superconducting by the proximity-effect (S'), but with a lower critical temperature, which would make it into a S'SS' device.  

The particular contact configuration, shown in Fig.~\ref{fig:3Dview} has been developed at Moscow State University of Education (MSUE) \cite{Tretyakov2011,Shcherbatenko2016}. In general, different choices have been made at different laboratories, mostly based on trial and error, with little conceptual guidance. The primary goal was to minimize a possible series resistance. The best practical devices \cite{Baselmans2004, Hajenius2006, Tarun2008, Zhang2014} have used instead of the Au-Cr-Au layer a NbTiN-Au or a Nb-Au layer on top of an Ar-cleaned NbN film. A similar strategy has been implemented at Cologne \cite{MunozJacobs2006,Puetz2011,Puetz2012, Puetz2015}. 

\begin{figure}[t]
\begin{center}
\includegraphics[width=\columnwidth]{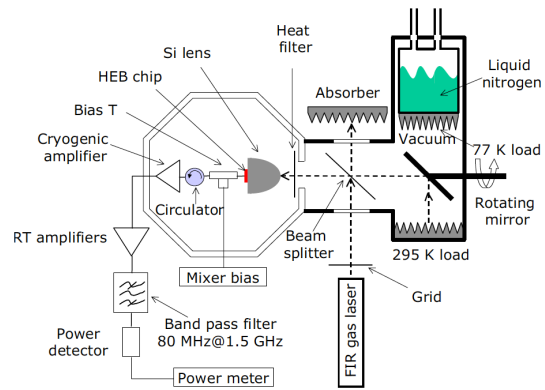}
\end{center}
\caption{\label{heteromixer} Standard set-up for a heterodyne mixer. The hot-electron bolometer mixer is mounted to a Si lens and indicated in red with HEBchip. A local oscillator is applied consisting of a far-infrared gas laser, but it could also be a quantumcascade laser (QCL). As a signal a black body source is applied as a 77 K load from liquid nitrogen and a room temperature load at 295 K. {(Reproduced with permission from the authors \cite{zhang2010})}}
\end{figure}

The superconducting hot-electron bolometer mixers emerged \cite{Gershenzon1990}, using niobium, in analogy to semiconductor bolometers. The initial concept for semiconductor hot-electron bolometers, proposed theoretically \cite{Rollin1961} and  experimentally realised in InSb by Arams et al \cite{Arams1966}. In all cases a uniform enhancement of the electron temperature is assumed, higher than the lattice-temperature or the phonon-temperature.  The mixing-process itself shows up in the electron temperature. {(For semiconductors, the use of a thermal electron distribution at an elevated temperature is an accurate description if the energy of the absorbed photons is not too high. As we will show, for superconductors the non-thermal nature of the distribution-function needs to be taken into account.)} Because the absorbed power by the electrons is quadratically dependent on the electric field the electron temperature will by the absorption of two signals with a different frequency (Fig.\ref{heteromixer}) develop a component at the difference frequency, the intermediate or IF frequency, $\omega_{IF}$. If the resistivity can follow this modulation of the electron temperature, also the resistivity will be modulated with that same frequency and a voltage signal at the IF-frequency will be measured. For semiconductors the resistivity is temperature dependent due the increase in carrier-density in low-gap semiconductors and due to a change in mobility. An important quantity, the upper limit for the IF-frequency is,  for a uniform temperature, determined by the energy relaxation rate from the electron system to the phonon system which is parametrised by the electron-phonon relaxation-time $\tau_{ep}$. The most successful semiconductor bolometer mixers were based on InSb, which had a rather long  $\tau_{ep}$, and hence a rather low maximum for $\omega_{IF}$. The major breakthrough  \cite{Gershenzon1990} was the discovery that superconducting thin films, which were known to have a fast electron-phonon relaxation time, could be used for mixing and provided a much higher maximum for the bandwidth of  $\omega_{IF}$. The experiments were initially carried out with niobium, Nb, and with 
YBa$_2$Cu$_3$O$_{7-\delta}$ with a modest bandwidth of 40 MHz.  These results were soon followed by results with NbN \cite{Goltsman1991} leading to a bandwidth of about 1 GHz. 

Initially, the strongest driving force for improvement of the mixer-performance was the IF bandwidth. Given the focus on the dependence on electron-phonon relaxation it meant selecting appropriate superconducting materials, which led early on to the move from niobium to niobium-nitride. However, in that process a very interesting proposition \cite{Prober1993} was to take advantage of the electrical contacts and use them as equilibrium reservoirs for rapid out-diffusion of the hot electrons of the superconductor. With a diffusion time \cite{Karasik1996b,Burke1999} given by $\tau_D=L^2/{\pi^2D}$ with $L$ the length of the device and $D$ the diffusion constant, this would provide an IF bandwidth of 4 GHz, easily outperforming the electron-phonon relaxation time in the initially used material Nb. The technological challenge would be to use advanced electron beam lithography to make short devices, while at the same time maintaining a well-matched impedance and use a high diffusivity superconductor. Hence, the desire to improve the IF bandwidth provided an additional reason for research on the contacts, but in this case with high diffusivity superconductors with a low normal state resistivity. In some laboratories \cite{Ekstrom1995,Skalare1995, Burke1996, WilmsFloet1999, Siddiqi2002} elemental superconductors such as niobium and aluminium have been used with interesting results for the physical processes. (In addition some work has been done on NbC  \cite{Karasik1996b, Ilin1998}.

These competing scenarios have, for about a decade, led to a classification of superconducting hot-electron bolometer mixers into \emph{lattice-cooled} (or \emph{phonon-cooled}) on the one hand and \emph{diffusion-cooled} on the other hand. The lattice-cooled devices had an IF bandwidth which was limited by the materials due to electron-phonon relaxation, whereas the diffusion-cooled devices were expected to be limited by the length of the device. As is clear from Fig.~\ref{fig:3Dview} the most successful material, NbN, emanating from the lattice-cooled development path has become embedded in a geometrically short structure such as proposed for diffusion-cooled devices. Therefore, the important distinction from the past has become obsolete, both phonon-cooling as well as diffusion-cooling play a role, at least in setting up the temperature profile. Nevertheless, the question remains: what limits the IF bandwidth in devices such as shown in Fig.~\ref{fig:3Dview}? This review reaches the conclusion that in practice the devices are, for their IF bandwidth, limited by phonon-cooling, because diffusion is blocked for the relevant electrons. These electrons are trapped in a potential well formed by the superconducting gap. Therefore, to improve the IF bandwidth, it is more beneficial, to search for other materials that can replace NbN. Interesting progress along this line has been made recently with magnesium diboride (MgB$_2$)\cite{Cunnane2015, NovoselovAPL2017}. A promising noise bandwidth of up to 11 GHz, combined with a higher operating temperature, has been reported. As we will show understanding the role of the contacts continues to be a relevant problem.

\section{Hot electrons and energy relaxation in metals and semiconductors}
 
\subsection{Uniform heating in normal metals and semiconductors}


The use of hot electrons {in semiconductors} for heterodyne mixing has been proposed theoretically \cite{Rollin1961} and demonstrated experimentally in InSb \cite{Arams1966}, exploiting the temperature dependence of the conductivity  at cryogenic temperatures. This increase in conductivity by radiation is not due to the creation of extra carriers, the usual photo-conductivity in semiconductors, but rather due to the energy-dependence of the mobility in low carrier density systems. An increase in temperature then leads to an increase in mobility, because of the increase in average energy. This \emph{photoconductivity of the second kind} \cite{Kogan1963}, benefits from cryogenic temperatures and, as suggested by Rollin,  is particularly interesting for mixing experiments. A very important driving force has been the discovery of the cosmic microwave background and the subsequent drive to obtain spectroscopic information in the hundreds of GHz range \cite{PenziasBurrus1973}. The first practical mixing device using hot electrons at frequencies around 100 GHz was presented for astronomical observations in the early 70-ies  \cite{PhillipsJefferts1973}.  

The main difference with this early work on semiconducting bolometers and the superconducting bolometers is in the temperature-dependence and the nature of the resistivity. The resistivity in semiconductors is, at low temperatures,  due to elastic backscattering taking into account their distribution over the energies. The dominant resistivity in superconductors is uniquely different because it is due to time-dependent changes of the macroscopic quantum phase. They have in common that they depend on the electron temperature, but the source of the resistivity and its temperature dependence is different. Meanwhile, the length of the devices has become short, bringing them into the mesoscopic transport regime.


\subsection{{\label{sec:pothier}}Hot electrons in mesoscopic normal metal devices}

Electronic transport in short devices was first considered for metallic pointcontacts \cite{Sharvin1965}. It was pointed out by Sharvin that two metallic reservoirs, connected by a small orifice with a size smaller than the mean free path for elastic scattering, has a linear Ohmic resistance, despite of the fact that there is no back-scattering. In his analysis he assumed an electron-reservoir on the right at a voltage $V$, and on the left on ground. Then the right-moving electrons originate from an electronic reservoir, characterized by a Fermi-Dirac distribution $f_0(E)$, whereas the left-movers originate at a reservoir with a distribution $f_0(E+eV)$. The difference between the left-movers and the right-movers is the current, which is given by the area of the orifice and the free electron parameters. The fundamental assumption is that electrons passing through the orifice conserve their energy. Their extra energy is shared with the bath by the interactions occurring in the electronic reservoirs. 

This conceptual framework has evolved further by including that the electrons also conserve their quantum-phase.  which led to an analysis of the conductance through a Sharvin pointcontact in terms of a quantum transport \cite{BeenakkerVanHouten1991}, analogous to a tunnel-junction. Since elastic scattering does not change the phase-coherence it includes quantum-coherent transport in the presence of elastic scattering, which is also energy-conserving. As before, energy-relaxation takes place in the equilibrium reservoirs. Various quantum corrections to the conventional Drude-transport have been identified. 

For superconducting hot-electron bolometers these normal metal quantum corrections are irrelevant because superconductivity is a much more dramatic change in conductivity. Nevertheless, the conceptual framework introduced by Sharvin, with its emphasis on energy-conserving transport processes, which connect equilibrium electron-reservoirs, is very relevant.  The main reason is that the properties of a superconductor are in many ways dependent on the distribution-function. The term hot-electron bolometer already implies that the distribution-function $f(E)$ is at least a non-equilibrium distribution function at an elevated electron temperature $T_e$. We will show below that in practice hot-electron bolometers should be described with a much more complicated dependence on the distribution-function than just its 'hotness'. Therefore, it is instructive to look first at voltage-biased mesoscopic normal wires \cite{Pothier1997}. 

Consider a narrow wire with diffusive scattering of electrons, connected to bulk contact reservoirs, labeled 1 and 2,  at different voltage $V$. In these reservoirs, for a given bath temperature $T_b$,  the energy distribution function is equal to the Fermi-Dirac distribution: $f(E)=[1+e^{({E+eU_{1,2}})/{k_BT_b}}]^{-1}$, with the voltage difference $V=U_1-U_2$. Along the length of the wire, equal to $L$, we assume a coordinate $X$. What is the energy distribution, $f(x,E)$ of the 'hot' electrons in the wire for a given voltage difference $V$? The answer is \cite{Pothier1997}: 

\begin{equation}
\label{eq:twostep}
f(x,E)=(1-x)f(1,E)+xf(2,E)
\end{equation}  
with $x$ defined by $X=xL$. This solution is the result of the diffusion equation:
\begin{equation}
\label{eq:diffeq}
\frac{1}{\tau_D}\frac{\partial^2 f(x,E)}{\partial x^2}+I_{coll}(x, E, \{f\})=0
\end{equation}  

with $I_{coll}(x, E, \{f\})$ the collision-integral which takes care of electron-electron and/or electron-phonon energy exchange. It is taken, for the solution given in Eq.~\ref{eq:twostep}, to be zero, meaning that over the length of the wire there is no energy-exchange between the electrons, nor with the phonon-bath. Transport along the wire is taken to be diffusive but energy-conserving. {In other words, the wire is shorter than the electron-electron and the electron-phonon relaxation length.}  This analysis is applied to an experimental study of copper wires, of which the length is varied \cite{Pothier1997}. Along the wires the local distribution-function $f(x,E)$ can be determined from the conductance of superconducting tunnel-junctions. They find for short wires indeed a \emph{non-thermal} two-step distribution function, which for longer wires gets more rounded due to electron-electron relaxation (the electron-phonon interaction in copper is much weaker).

This non-thermal distribution evolves into a thermal distribution for longer wires due to the fact that energy gets exchanged between the electrons leading to a local Fermi-Dirac distribution with an effective temperature $T_{eff}(x)$ given by: 

\begin{equation}
\label{eq:thermal}
T_{eff}(x)=\sqrt{T_b^2+x(1-x)\frac{V^2}{L_0}}
\end{equation}
with $L_0=\frac{\pi^2}{3}(\frac{k_B}{e})^2$ the Lorenz number. This simple result reflects the fact that the thermal resistance and the electrical resistance are connected by the Wiedemann-Franz law. In the experiment \cite{Pothier1997} the experimentally relevant electron-electron time turned out to be influenced by remnant magnetic impurities \cite{Huard2005}. This analysis makes clear that only under certain conditions an enhanced electron temperature $T$ is adequate to characterize  the nonequlibrium state of the electron system. In many cases a non-thermal distribution characterized by $f(E)\neq f_{FD}(E)$ is more appropriate. As we will see this is in particular important for superconductors because of the complex dependence of the superconducting properties on deviations from $f_{0}(E)$. In particular, because superconducting hot-electron bolometers convert the distribution-function into the resistive properties of a superconductor, which ultimately generate the observed signal at the intermediate frequency.   

Superconducting hot-electron bolometers are brought to their operating point by a dc bias \emph{and} and by the application of a signal from a local oscillator (Fig. \ref{heteromixer}). The effect on the distribution-function by the dc bias is in principle captured in the framework of the previous paragraph \cite{Pothier1997}. The signal from the local oscillator contributes also to the non-equilibrium distribution. In principle, one approach to model this would be to replace in the previous analysis $V$ by a time-dependent voltage as is commonly done for tunnel-junctions, in the concept  of photon-assisted tunnelling \cite{TienGordon1963,Tucker1985}. This approach leads to an even more complicated non-equilibrium distribution than given by 
Eq.~\ref{eq:twostep}. Such an analysis has been implemented theoretically \cite{Shytov2005, Dikken}, but without an experimental evaluation. Such a voltage-driven radiation-source approach will depend on the electron-electron interaction time over the length of the wire, in combination with the frequency dependence. The relevant parameters for various materials are provided in Table \ref{tab:Table3}. Evidently, for the commonly used material NbN the electron-electron interaction time is very short. In order to proceed we take as a starting point that the most interesting regime for the application of the hot-electron bolometers is beyond 1 THz. In addition, there is a universal experimental preference for NbN. And finally, the used NbN has a critical temperature  in the order of 10 K, meaning a superconducting energy gap in the order of 3 mV. As we will show, focusing on NbN, we need for the analysis of the DC properties of hot-electron bolometers the approach followed by Pothier et al.. However, for the properties under operating conditions it is sufficient to assume that for $\hbar\omega>>2\Delta$ and $\omega\tau_{ee}\sim 1$ a uniform absorption of the LO-signal is appropriate, which contributes to an elevated electron temperature.      

\begin{figure}[t]
\begin{center}
\includegraphics[width=1\linewidth]{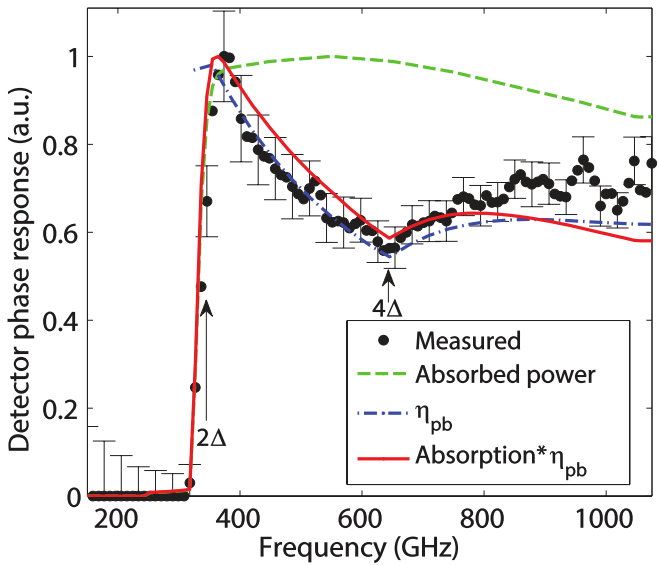}
\end{center}
\caption{\label{fig:Figure2} The measured THz absorption of a tantalum MKID-detector (\cite{DeVisser2015}) as a
function of the frequency (normalised to
one). The green dashed line represents
a calculation of the power absorption of the superconducting transmission
line, including the antenna efficiency. The blue dashed-dotted line is a simulation
of the pair-breaking efficiency (not normalised) that arises due to the
different quasiparticle distributions at different excitation frequencies. The
red line combines the two effects (red and green lines both normalised
to one). Below 350 GHz there is no absorption because $\hbar\omega<2\Delta$. Above 350 GHz the absorption rises quickly with features which depend on the energy-dependence of the quasiparticle and the phonon-system. {(Reproduced with permission from the authors \cite{DeVisser2015})}}
\end{figure}

Therefore, we will assume that the power from the radiation is  uniformly absorbed by the electron system $P_{rf}$. This power which enhances the energy of the electron system will be modeled by an increase in electron temperature. This is commonly done for superconductors with a photon-energy $\hbar\omega$ higher than the superconducting energy gap $2\Delta$, needed to break Cooper-pairs. However, in superconductors the energy-relaxation is controlled by the recombination of quasiparticles to Cooper-pairs, which leads to the emission of $2\Delta$ phonons and hence a non-equilibrium distribution of phonons as well. Already in early work \cite{RothwarfTaylor1967},  it was found that this recombination-time is limited by the phonon-escape rate out of the thin metal film. Additionally, it is important to understand that, unlike a semiconductor, for a superconductor an increase in temperature does not lead to an increase in resistance in a superconductor. It leads to an increase in the density of quasiparticles and a decrease in the energy gap, but the resistance stays zero, unless one is very close to the critical temperature. Nevertheless, it would be possible to monitor the changes in superconducting properties by measuring its kinetic inductance at GHz frequencies.  In recent work  \cite{DeVisser2015} on microwave kinetic inductance detectors, made from tantalum, the absorption of energy in the THz range around the $2\Delta$ energy has been mapped (Fig.~\ref{fig:Figure2}). Obviously below 350 GHz there is no absorption. At the threshold it sharply rises followed by a curve which also shows the influence of the phonon-bath at 2 times $2\Delta$.        

It illustrates that an important ingredient in the absorption of radiation by the electron-system, leading to an elevated temperature for the electrons, are the phonons.  The recombination of quasi-particles to Cooper-pairs needs to be broken down in a two-step process. Photons with energy higher than the energy gap break Cooper-pairs leading to an excess of quasiparticles, which can in principle also be interpreted as a higher effective electron temperature. The quasiparticles can relax to a lower energy level by scattering processes, called phonon-cooling, including the emission of phonons. The quasiparticles can also recombine to form Cooper pairs with the emission of so-called $2\Delta$-phonons 
(Fig.~\ref{fig:Figure2}). This will lead to a nonequilibrium phonon-bath and the phonons need to equilibrate by interaction with the substrate phonons. Since the substrate and the superconducting metal have different sound velocities and different atomic densities the phonons do experience a barrier at the interface, which is analogous to the acoustic mismatch resistance known as the Kapitza resistance. These processes have been analyzed  \cite{Kaplan1976} and applied to the temporal response of the optical absorption \cite{Perrin1982,PerrinVanneste1983}. To minimize the effect of the phonons one increases the escape rate by making the films as thin as possible, which for niobium-nitride has led to an optimal thickness of a about 4 nanometer.

\subsection{\label{sec:hotsuper}Hot-electrons and non-equilibrium in superconductors}
Compared to normal metals and semiconductors the subject of hot electrons and superconductors is quite complicated. A superconductor has in principle a resistivity equal to zero. If the electrons are 'hot', i.e. an increase in electron temperature the Fermi-Dirac distribution will be described by $T_e>T_{bath}$. As a consequence the energy gap $\Delta$ will gradually decrease, as well as the critical pair-breaking current $j_c$ and the critical magnetic field $H_c$. Despite of all these changes the resistance remains zero. As a consequence the modulation of the electron temperature at the IF frequency will not be observable. Only when the superconductor shows signs of resistance an observable signal emerges. In Section \ref{sec:resistsuper} we will discuss the potential sources for resistance in a superconductor, which can respond to a modulation of the electron temperature. First, we will discuss the nature of an elevated electron temperature in a superconductor. 

The properties in a superconductor are controlled by the distribution-function $f(E)$, which is in equilibrium given by the Fermi-Dirac distribution $f_0=[1+\exp{\mu/k_B T}]^{-1}$, with $\mu$ the electrochemical potential, usually the Fermi-energy, $k_B$ Boltzmann's constant and  $T$ the electron temperature, usually equal to the phonon-temperature. Superconductivity is a state of matter in which electrons, known from normal metals, condense into a new state in which electrons form pairs, Cooper-pairs, with opposite momenta and opposite spin. At $T=0$ all electrons are condensed into this state of pairs. At a finite temperature a few pairs break up and unbound quasiparticles are created. This process is controlled by the Fermi-Dirac distribution, $f_0$. The pairs are bound with an energy $\Delta$. A few electrons will have a thermal energy higher than $\Delta$, which causes the emergence of excitations. The excitations in a superconductor are Bogoliubov quasiparticles, which means that they are a mixture of electrons ($k>k_F$) and holes ($k<k_F$). The higher $k$ the stronger the quasiparticle resembles an electron, similarly the deeper in the Fermi-sea the stronger it resembles a hole. For $k=k_F$ the quasiparticles are equally hole and equally electron. This also means that their charge is equal to zero.         

In order to characterise the nonequilibrium nature of a superconductor it has been recognized \cite{Tinkham} that the deviation form the Fermi-Dirac distribution, $\delta f(E$), should be split into a symmetric part and an asymmetric part around $k_F$. For the symmetric part of $\delta f$ both branches of the excitation spectrum deviate from equilibrium similarly, analogous to an increase in temperature and the absorption of radiation. In other words this is the 'hot electrons' part, although deviations do not have to be thermal and neither do they have to be only positive, it can also be negative, i.e. symmetric depopulation of states. For convenience we will call this mode of nonequilibrium the \emph{energy mode}. The asymmetric mode of nonequilibrium concerns cases in which the branch for $k>k_F$ is differently populated than the branch with $k<k_F$. It occurs when the superconductor is fed by a current from a non-superconducting reservoir, such as a normal metal. If normal charge is fed into a superconductor the electron-like branch is more strongly populated than the hole-like branch. As a consequence, compared to an equilibrium superconductor there is an excess charge in the quasiparticle system, which over a certain distance will convert into an equilibrium situation. This mode of nonequilibrium is called the \emph{charge-mode}. Interestingly, it leads to a static electric field inside the superconductor, which drives the excess quasiparticle charge and can be measured as a dc voltage in a superconductor. Since a hot-electron bolometer is a device which absorbs radiation as well as being coupled to normal electrodes, or electrodes with a much lower critical temperature, both types of nonequilibrium play a role. The term 'hot-electrons' suggests erroneously that only the energy-mode of nonequilibrium plays a role. We will show that the charge-mode plays a role as well, and it would in principle be better to call them \emph{superconducting nonequilibrium mixers}. In this respect superconducting hot-electron mixers differ fundamentally from semiconductor hot-electron mixers.              

In early descriptions of superconducting hot-electron bolometer mixers they were understood as having a temperature dependent resistivity. In semiconductors both the carrier concentration and the mobility can vary in temperature according. For semiconducting THz mixers the temperature dependence of the mobility is used. In contrast, the conductivity of a normal metal from which the superconducting state emerges is temperature-independent. The interesting property of a superconductor is the transition from the normal state to the zero-voltage state, which is called the resistive transition.  A typical resistive transition, shown in Fig.~\ref{fig:SITNbN2013} is parametrised by an empirical formula:

\begin{equation}\label{empirical rho(T)}
\rho(T)=\frac{\rho_0}{1+e^{-\frac{(T-T_c)}{\Delta T_c}}}
\end{equation}  
with $\Delta T_c$ the width of the transition, $T_c$ the midpoint of the transition and $\rho_0$ the resistivity in the normal state. The temperature $T$ appearing here should be the electron temperature, meaning that in Fig.~\ref{fig:SITNbN2013} the horizontal axis indicates the electron temperature $T_e$. In principle it is assumed that Eq.~\ref{empirical rho(T)} is an adequate representation of the temperature dependence of the resistive properties of the superconducting film.   

Analogous to the semiconductor case we now assume that we can bias the superconductor with a current $I_0$, using wires connected to the superconductor, which do not physically effect the electron temperature. Furthermore we assume that we can operate the device at an electron temperature $T_e$ different from the bath temperature $T_b$ and which brings the superconducting film in the strongly temperature dependent regime around $T_c$. With these assumptions the superconductor can be analyzed in exactly the same manner as the semiconductor hot-electron bolometer. The only difference is that the expression for the temperature dependence of the resistance is different. In practice however this recipe is highly deceptive and the use of it has delayed the development of a usable framework for understanding and optimising superconducting hot-electron bolometers. The missing ingredient is a discussion of the question how a superconductor, which by definition is supposed to have $R=0$ can nevertheless develop resistance.

\section{\label{sec:resistsuper}Resistive superconducting properties}

\subsection{\label{sec:SIT}Resistivity in the normal state and superconductivity: superconductor-insulator transition}
The electromagnetic properties of BCS superconductors have been theoretically calculated by Mattis and Bardeen \cite{MattisBardeen1958} and generalised by Nam \cite{Nam1967a,Nam1967b,Nam1970}. It expresses the complex conductivity $\sigma=\sigma_1+i\sigma_2$ normalised to the normal state conductivity $\sigma_n$. This normalisation takes into account that s-wave BCS superconductivity is not influenced by elastic scattering, an insight known as \emph{Anderson's theorem} \cite{Anderson1959}. The great advantage is that superconducting films can tolerate a lot of variation in resistivity, while maintaining good superconducting properties. 

For optimal coupling of electromagnetic radiation in the THz range one aims for normal state resistances that match to free space and other electrically relevant components, which means about 75 $\Omega$. In practice it has led to the use of, in particular niobiumnitride (NbN) films, with a thickness of about 4 nm, a critical temperature of about 10 K, and an area of 4 $\mu$m by 0.4 $\mu$m. These films have a resistance per unit area, $R_{\square}$ of about 1000 $\Omega$ or a normal state resistivity of 500 $\mu\Omega$cm, which implies an elastic mean free path in the order of the interatomic distance. This puts the material in the regime of \emph{strongly disordered superconductors} for which  \emph{Anderson's theorem} breaks down. 

\begin{figure}[t]
\begin{center}
\includegraphics[width=1.0\columnwidth]{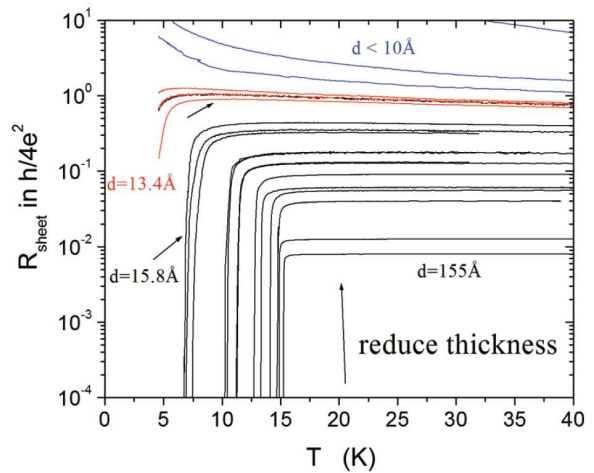}
\end{center}
\caption{\label{fig:SITNbN2013} Resistive transitions for NbN films of different thicknesses, scaled to the quantum unit of resistance $h/{4e^2}=6.45~k\Omega$. Films used for hot-electron bolometer mixers usually have a normalised value of $R_{sheet}=R_\square \approx0.2$ close to the superconductor-insulator transition. {(Reproduced with permission from the authors \cite{Yong2013})}} 
\end{figure}

The strongly disordered superconductors are being actively studied in the context of \emph{superconductor-insulator transitions} (SIT). The fundamental question is how superconductivity gets destroyed by increasing disorder, increasing impurity-scattering. Similarly, one can study how the metallic state can be destroyed by increasing disorder which has led to the concept of localisation of wave-functions in contrast to extended states like in a genuine metal, as well as to the increased effect of electron-electron interaction in contrast to free electrons. Naturally, superconductivity is in competition with increased Coulomb repulsion and a tendency to localisation, i.e. insulating behaviour. This is an active field of research, which includes the materials NbN, TaN, TiN, NbTiN and InO$_x$, and one which does not provide easy answers quickly. One of the fundamental questions is whether the amplitude of the order parameter decreases with increasing disorder or does the phase-correlation of the order parameter break down, maintaining localized Cooper pairs. The current wisdom is that in realistic systems both occur intermixed. 

\begin{figure}[t]
\begin{center}
\includegraphics[width=1.0\columnwidth]{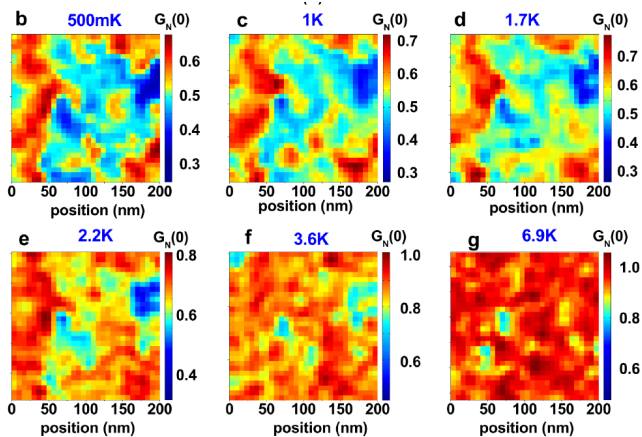}
\end{center}
\caption{\label{fig:NbNfilm} Scanning tunnelling pictures of a NbN film with a resistively measured $T_c$ of 2.7 K and a pseudogap $T_c$ of 7.2 K. The colorscale indicates the depth of the minimum in tunnelling conductivity signalling the opening of the energy gap. {(Reproduced with permission from the authors \cite{Pratap2013})}} 
\end{figure}

An important and significant result of local tunnelling experiments  on thin films of NbN and TiN is that spatial fluctuations of the superconducting energy gap are found 
(Fig.~\ref{fig:NbNfilm}). These spatial fluctuations arise from the competition between localisation and superconductivity, perhaps amplified by metallurgical imperfections. Whatever its origin, it is unavoidable that NbN films with a resistivity in the range $500~\mu\Omega cm$ will not have uniform superconducting properties. Also small pieces of a large superconducting film my have different properties dependent on the correlation length of the spatial fluctuations. This will fundamentally limit the fabrication yield of the devices. 

Recently, interesting results have been obtained with a new BCS superconductor, magnesium diboride, MgB$_2$. At present no systematic study has been made of this material within the context of the superconductor-insulator transition. The requirement of impedance matching will also demand high resistivity material, which could partially be achieved by a lower carrier density. 

\subsection{Resistivity of superconducting wires: phase slips}
Superconducting hot-electron bolometers use, in principle,  the sensitivity of the resistance of a superconductor to small changes in temperature. This may sound confusing, because the name of the superconducting state, superconductivity, refers directly to one of the unique properties of a superconductor, zero resistance. How can a superconductor have resistance? The source of resistance in a superconductor, and its temperature dependence, is unrelated to the resistance in the normal state. The latter is in the relevant temperature range roughly independent of temperature. The superconductor has resistance due to time-dependent changes in the macroscopic quantum phase. In principle a supercurrent is driven by a gradient in this phase, which can be uniform in a wire with a small cross-section.  As understood from the Josephson-effect if the phase-changes in time, $d\phi/dt$, it will be accompanied by a voltage $2eV/\hbar$. If the phase slips by $2\pi$ a voltage spike occurs, without changing the superconducting state. With increasing temperature the likelihood of phase-slip events increases leading to an increase in the number of voltage-spikes, which on average is observed as a finite voltage. This is the cause of the resistive transition in a one-dimensional wire. 

The original concept was proposed by Little\cite{Little1967} triggered by considering superconductivity in long molecules. The standard analysis of so-called thermally assisted phase slips has been summarized in textbooks  \cite{Tinkham}, which for the current-voltage characteristic leads to: 

\begin{equation}\label{TAPP}
V=\frac{\hbar\Omega}{e}e^{-F_0/{kT}}\sinh\left(\frac{hI}{4ekT}\right)
\end{equation} 

which for small currents reduces to 
 \begin{equation}\label{TAPP_R}
R=\frac{V}{I}=\frac{\pi\hbar^2\Omega}{2e^2kT}e^{-F_0/{kT}}
\end{equation} 
These results have been very well tested in tin whiskers. The theory has been reanalysed in the context of the study of quantum phase slips \cite{Golubev2008} with a full review of the topic including experiments \cite{Arutyunov2008}. As illustrated in Fig.~\ref{fig:3Dview} practical devices are rather wide, in particular compered to the coherence length in niobium-nitride cf. Table \ref{tab:Table3}. The concept of phase-slip assumes a one-dimensional object in which the macroscopic quantum phase can not change in a lateral direction. Therefore we assume that for practical hot-electron bolometers the resistive transition is not controlled by thermally assisted phase slip processes but by the 2-dimensional equivalent, which is connected to the Berezinskii-Kosterlitz-Thouless phase transition.

\begin{figure}[t]
\begin{center}
\includegraphics[width=0.6\columnwidth]{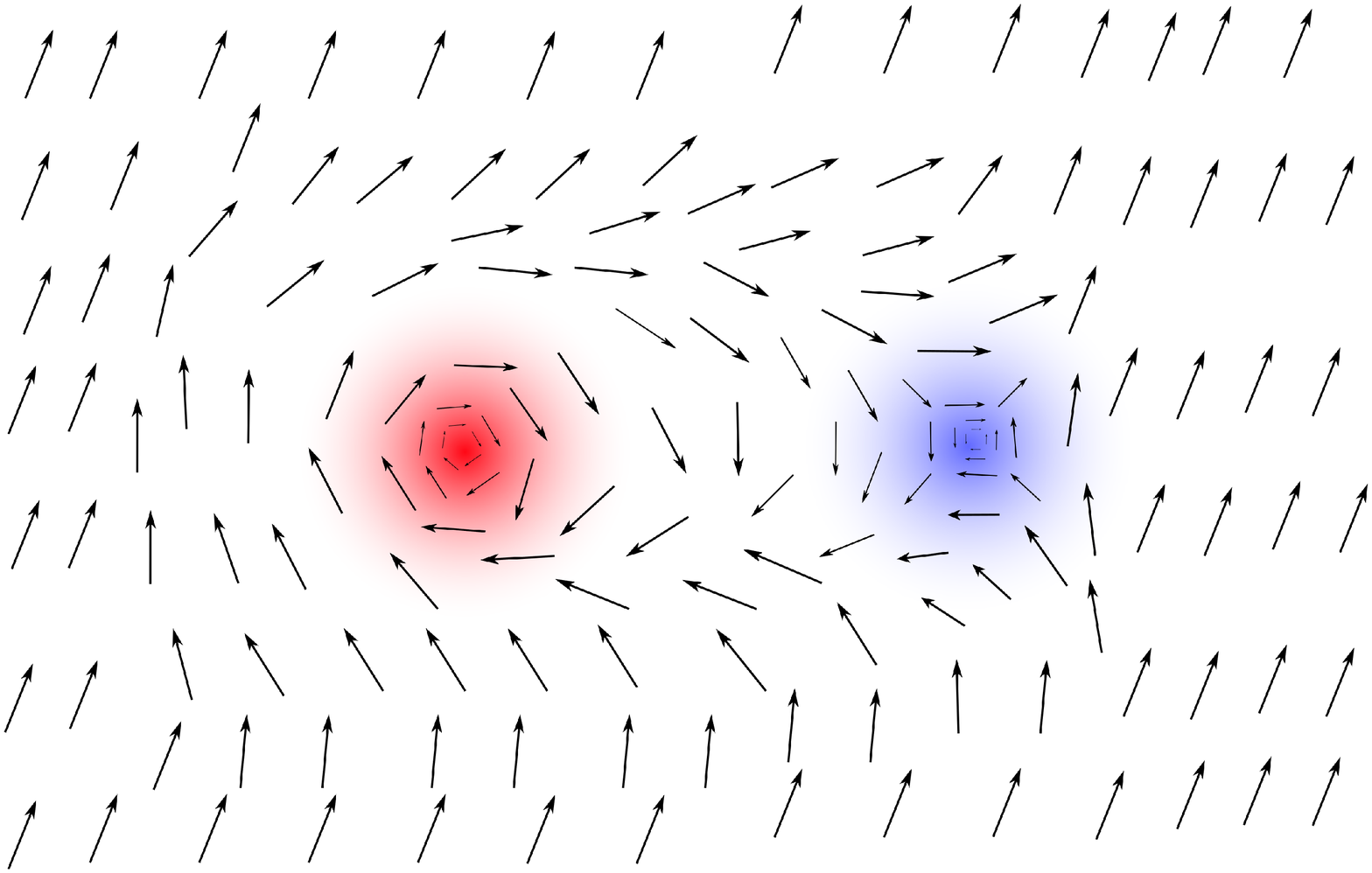}
\end{center}
\caption{\label{fig:VAV} Sketch of a vortex-anti-vortex pair in a uniform superconducting film. Arrows indicate an angle, being the direction of the phase of the superconductor at those points, the length the absolute value of the order parameter. The blue and the red vortex have opposite rotations of $2\pi$. Each vortex has a core of size $\xi$. The phase-change is accompanied by a circulating supercurrent with a logarithmic decay controlled by the Pearl penetration depth, $\lambda_\perp$. {(Reproduced with permission from the author \cite{Beekman2011})}} 
\end{figure}

\subsection{\label{sec:BKT}Resistivity of 2-dimensional superconducting films: Berezinskii-Kosterlitz-Thouless phase-transition}
In the 70-ies the superconductivity of two-dimensional films was addressed theoretically \cite{Berezinskii1971,Berezinskii1972,KosterlitzThouless1973,Kosterlitz1974}. It showed the breakdown  of long range order in physical systems with a continuous order parameter such as superconductors. The relevance of these general theoretical considerations became apparent for thin film superconductors through the work of Beasley and co-workers \cite{BeasleyPRL1979} and Halperin and Nelson \cite{HalperinNelson1979}. The fundamental statement is that two-dimensional superconductors in equilibrium, which are normally characterized by an order parameter $\psi$ with a uniform phase $\phi$, should be viewed differently. In the ground state vortices  and anti-vortices are bound together in pairs (Fig. \ref{fig:VAV}), which at higher temperatures dissociate leading to a plasma of free vortices with opposite signs. This dissociation sets in at the BKT temperature $T_c^{{BKT}}$, which in general is close to but below the BCS mean-field theory value of the critical temperature: $T_{c0}$.    

As demonstrated for thin films \cite{BeasleyPRL1979} to bring these superconductors in the regime where  $T_c^{{BKT}}<<T_{c0}$ one needs films with a very large perpendicular penetration depth, also called Pearl length, which is given by: 
\begin{equation}
\lambda_\perp=\frac{\lambda^2}{d}=\frac{\lambda_L^2(0)}{d}\left(\frac{\xi_0}{l}\right)\left[\frac{\Delta(T)}{\Delta(0)}\tanh\left(\frac{\beta\Delta(T)}{2}\right)\right]^{-1}
\end{equation}
with $d$ the thickness of the film, $\lambda_L$ the London penetration depth, $\xi_0$ the BCS coherence length, $l$ the elastic mean free path, $\Delta$ the energy gap and $\beta=1/(k_BT)$. In order to bring the material in the desired regime one prefers thin films with a short mean free path.  For realistic parameters this amounts to values in the hundreds of $\mu$m range. By including relevant numbers one finds that $T_c^{{BKT}}$ is expected to be given by: 

\begin{equation}
\frac{T_{c}^{BKT}}{T_{c0}}=\left( 1+0.173(\times2)\frac{R}{R_Q}\right)^{-1}
\label{eq:Tbkt}
\end{equation}
with $R_Q$ the quantum unit of resistance per square of $2h/{e^2}=12.9~k\Omega$. Obviously, a larger sheet resistance leads to a lower value of  $T_c^{{BKT}}$ compared to $T_{c0}$. 

The resistive superconducting properties appear when there are free vortices which can move under the influence of a current due to a Lorentz force perpendicular to the direction of the current. The resistive properties of the superconductor \cite{HalperinNelson1979} arising in superconductor above $T_c^{{BKT}}$ leads to the expression of the conductivity of:

\begin{equation}
\label{eq:nelsonhalperin}
\sigma_s\approx 0.37\frac{1}{b}\sinh^2 \left[b\frac{T_{c0}-T_c^{BKT}}{T-T_c^{BKT}}\right]^{1/2}
\end{equation}
with $b$ a numerical constant of order unity. This expression leads to a rapid rise of the resistivity beyond $T_c^{{BKT}}$. 

From Eq.~\ref{eq:Tbkt} it is apparent that a sheet resistance will lead to a value of $T_c^{{BKT}}$ well below $T_{c0}$. However, in Section \ref{sec:SIT} it has been pointed out that an increase in sheet resistance leads also to non-uniform properties of the amplitude of the order parameter, which needs to be taken into account in a comparison between theory and experiment. Additionally, in practice superconducting films are finite, whereas the theory assumes an infinitely large system. Both aspects have been taken into account \cite{Benfatto2009}  leading to a fairly good agreement between theory and experiment in a number of cases such as shown for NbN films \cite{KamlapureAPL2010, Yong2013}. 

The NbN films routinely used for hot-electron bolometers {are in a range of Fig.~\ref{fig:SITNbN2013}) where the superconductor-insulator transition is approached.} It is therefore reasonable to expect that upon approaching a temperature where resistivity sets in,  this resistivity is due to the emergence of free vortices, which move under the influence of a current.   
   
\subsection{{\label{sec:conversion}}Charge conversion resistance at normal-metal-superconductor interfaces}

\begin{figure}[t]
\begin{center}
\includegraphics[width=1.0\columnwidth]{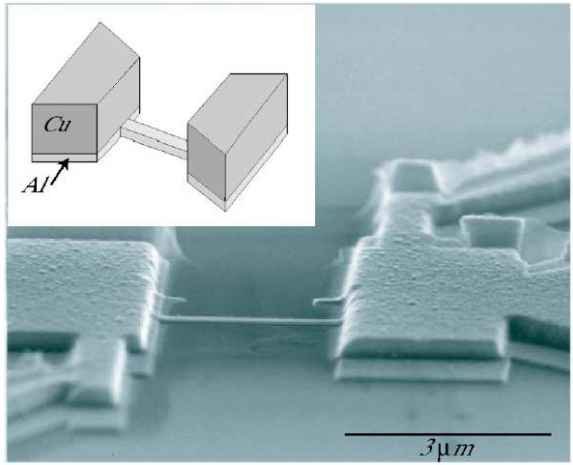}
\end{center}
\caption{\label{fig:NSNsample} Scanning electron microscope picture of a Cu-Al-Cu NSN device (slightly misaligned) 
\cite{Boogaard2004}, showing the coverage of the thin aluminum film with the thick Cu layer. The inset shows a 
schematic picture of an ideal device. {(Reproduced with permission from the authors \cite{Boogaard2004})}}
\end{figure}

\begin{figure}[t]
\begin{center}
\includegraphics[width=1.0\columnwidth]{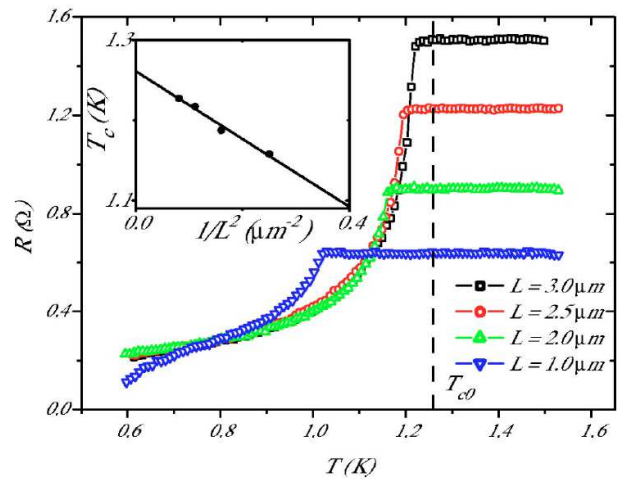}
\end{center}
\caption{\label{fig:Plateau} Measured R-T curves for four different NSN bridge lengths of the device shown in Fig.~\ref{fig:NSNsample}. 
The intrinsic $T_{c0}$ is indicated by the vertical dashed line. The inset shows the measured critical temperature of the wire 
versus $1/L^{2}$, which is used to determine $T_{c0}$ by letting $L\rightarrow \infty$. {(Reproduced with permission from the authors \cite{Boogaard2004})}} 
\end{figure}

The understanding of hot electrons and non-equilibrium distributions in normal metal wires connected to normal metal equilibrium reservoirs leads naturally to the question of a superconducting wire between two normal equilibrium reservoirs. Such an NSN system can serve as a model system to understand hot electron bolometer mixers. This system has been studied experimentally and theoretically in series of papers \cite{Boogaard2004, Keizer2006,Vercruyssen2012}.  Fig.~\ref{fig:NSNsample} shows a SEM-picture of an aluminium wire of a few micrometers long connected to copper contacts. Aluminium has a weak electron-phonon interaction and therefore a long inelastic scattering length. In addition, the electron-electron interaction is also weak. In this 2-point measurement one finds for the resistance for different lengths the set of curves shown in Fig.~\ref{fig:Plateau}. The material has an intrinsic critical temperature marked by $T_{c0}$. These resistive transitions measured for wires with different lengths show for shorter wires a lower value for the temperature where the resistance decreases. This is due to the lateral proximity-effect, for a shorter wire the influence of the normal state of the contacts is more strongly felt. Strikingly, one observes that all curves reach with decreasing temperature the same value of the resistance. It suggests that there is a contact resistance (the descent of the blue curve at lower temperatures is because the contacts experience superconducting correlations, which can be further ignored). This resistance is due to evanescent states of incident electrons which are converted into Cooper-pairs. It exists inside the superconductors over a length of about the coherence length.      

This charge-conversion resistance is not very well known. It occurs at interfaces between a normal metal and a superconductor and in cases when normal charge is injected into a superconductor for example in a tunnel-junction. In an analysis of experiments close to $T_c$, where $kT>>\Delta$ the majority of electrons have an energy high enough to enter the superconductor as quasiparticle charge. It was originally introduced as \emph{branch imbalance}, $Q$,  \cite{TinkhamPRL1972, TinkhamPRB1972} and subsequently labeled  \emph{charge imbalance},  $Q^*$ \cite{PethickSmith1979}. For an NIS tunneljunction only the fraction $F^*$ of the electrons with energy larger than $\Delta$ can tunnel into the superconductor creating the excess of charge and an accompanied difference in chemical potential between the quasiparticles $\mu_{qp}$ and the Cooper pairs $\mu_p$ becomes a measurable quantity. At an interface between a normal-metal and a superconductor interface it was assumed that only quasiparticles with energy $E$ larger than $\Delta$ would be able to enter the superconducting state, causing an excess of quasiparticles moving in S in the direction from N to S, creating an excess of quasiparticle charge, which had to relax to zero over a length of $\Lambda_{Q^*}=\sqrt{D\tau_{Q^*}}$, with $D$ the normal metal diffusion coefficient. The charge relaxation time, $\tau_{Q^*}=\tau_{inel}\sqrt{4kT_c/\pi\Delta}$,  is essentially controlled by the inelastic scattering $\tau_{inel}$ and a divergent factor dependent on the superconducting gap. 

The new element in the observation of Fig.~\ref{fig:Plateau} is that also at much lower temperatures where $kT<<\Delta$ a charge-conversion resistance is present. Previously, it was understood that when the process of Andreev reflection is present there is no interface resistance for charge, only for thermal conductivity.  In earlier experimental work on an NS-interface quantitatively it was found that no interface resistance is present in this temperature range  \cite{HsiangClarke1980}.  This was understood by the assumption that the fraction of the current, $1-F^*$,  for which the energy $E<<\Delta$ does not contribute to the resistance, because it would be Andreev-reflected. This insight was, erroneously, taken over in Section VII of the well-known BTK-theory \cite{BTK1982}. The error is in the lack of understanding of the contribution of the Andreev-process to the resistivity of the superconductor, which has become more urgent given the present-day nanoscale of device structures. The first experimental observation of such an unexpected resistance contribution from the superconductor was reported in 1974  \cite{Harding1974}. They reported for Cu/Pb$_x$Bi$_{1-x}$ a strong dependence of an excess resistance beyond the one expected for copper and depedent on the impurity scattering in  Pb$_x$Bi$_{1-x}$, which could be tuned by the Bi concentration.  However, the same observation was absent in ater experiments  \cite{HsiangClarke1980}. Two decades later new evidence was reported in niobium  \cite{GuPratt2002} in 2002, while making an attempt to study the dependence on spin polarisation with FS structures. 

The most systematic and convincing experimental results have been obtained by using the 2-point resistance of an aluminium wire with two thick and wide normal contacts as equilibrium reservoirs \cite{Boogaard2004}. The length of the aluminium wire was changed and a fixed value for the resistance in aluminium was found down to temperatures far below $T_c$. This was understood as due to the charge imbalance of evanescent states occurring over a length of about the coherence length. The results  \cite{Boogaard2004} have been compared with the Keldysh Greens function theory. It has been found that it corresponds to a resistance of a piece of superconductor with a length of about $\sqrt{{\pi}\xi_0 l/6}$, with $l$ the elastic scattering length and $\xi_0$ the BCS coherence length.  
\begin{figure}[t]
\begin{center}
\includegraphics[width=1.0\columnwidth]{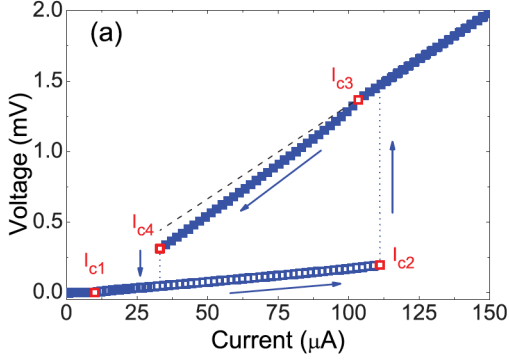}
\end{center}
\caption{\label{fig:NathanIV} The voltage $V_{12}$  of a 4-$\mu\mathrm{m}$-long wire as a function 
of bias current $I_{12}$, measured at 200 mK. We define four different regimes with 
boundaries labeled $I_{c1}\textrm{-}I_{c4}$, each characterized by a nearly constant differential 
resistance. The critical currents $I_{c2}$ and $I_{c4}$ are defined as the bias currents where 
the wire switches between the two hysteretic voltage branches. $I_{c1}$ and $I_{c3}$ are the 
transition points between the two different states of one branch. {(Reproduced with permission from the authors \cite{Vercruyssen2012})}}
\end{figure}
Subsequently the voltage-dependence was addressed in a theoretical paper \cite{Keizer2006}, in which the dependence of the superconducting properties on the distribution-function was taken into account, followed by experimental work \cite{Vercruyssen2012}.  

 \begin{figure}[t]
\begin{center}
\includegraphics[width=\columnwidth]{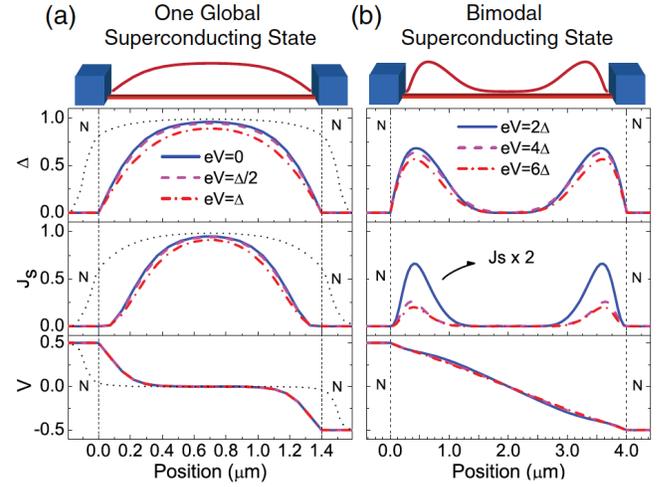}
\end{center}
\caption{\label{fig:NathanProfiles} (a) The complete wire is in a single superconducting state with order parameter $\Delta(x)$. 
Near the normal reservoirs, the condensate carries only a small fraction $J_{s}$ of the current as a supercurrent, 
which results in a resistance and a voltage drop at the ends of the wire, over roughly a coherence length. At the lowest 
temperatures, a small proximity effect can occur at the connection of the bilayer reservoirs to the wire (schematically 
illustrated by dotted black lines). (b) Two distinct superconducting domains at the ends of the wire are separated by a 
normal region in the center of the wire. Due to the small supercurrent, the voltage profile is almost equal to the normal state. {(Reproduced with permission from the authors \cite{Vercruyssen2012})}} 
\end{figure}

In the work of Vercruyssen et al the full current-voltage characteristic of a NSN structure (Fig. \ref{fig:NathanIV}) is traced in detail and interpreted using the Keldysh Green's function approach \cite{Keizer2006}, in which two kinds of nonequilibrium: transverse and longitudinal are dealt with on equal footing. This is based on a distribution function controlled in the manner first introduced for normal mesoscopic wires \cite{Pothier1997}. This work has demonstrated the following significant facts about NSN structures:
\begin{itemize}
\item the superconducting wire has a resistance approximately equal to normal state resistivity times the coherence length for dirty superconductors
\item the resistance is hardly effected by an increasing current (or voltage) indicating that the power is not controlling the resistance but rather the charge conversion
\item this resistance is terminated by what one would be tempted to call a critical current but what we believe is rather a critical voltage indicative of the power accumulated in the electron system, in analogy with a temperature rise leading to a quench of the superconducting state
\item{before this critical value the order parameter is expected to have a flat profile tapering off at the edges due to the proximity-effect}
\item {beyond the critical voltage the wire becomes completely normal}
\item with decreasing voltage-bias superconductivity starts to emerge at the edges, the coolest parts of the driven wire, and the order parameter has a camelback profile
\item the hysteresis curve is terminated when the camelback profile flips to the flat profile  
\end{itemize}
This scenario has been inferred from experimental work on aluminium wires supported by theoretical work on non-equilibrium superconductivity taking into account the subtle dependence of the superconducting state on the distribution-function, which goes beyond the dependence on temperature. It underlines that \emph{hot electrons} is in need of a more sharp definition  for experiments on superconductors under nonequilibrium conditions.

\section{{\label{sec:barends}}DC characterization of superconducting hot-electron bolometers}

In principle a hot-electron bolometer device is a NSN configuration. The two N-parts are equilibrium reservoirs at the bath temperature $T_b$, where electron and phonons are in equilibrium. For the electrons in these reservoirs it means that their distribution over the energies  is characterized by the Fermi-Dirac distribution. The S-part, made of NbN, has material parameters shown in Table \ref{tab:Table1}.  In practice the two N electrodes can also be a superconductor, S', intrinsically or due to the proximity-effect, with a lower $T_c$. As we will show below to understand the DC behavior of these devices, when no external radiation is applied, we need to treat the devices as mesoscopic structures in which the current flows in response to an applied voltage difference. This is easily envisioned for a NSN device, but is more subtle for a S'SS' device. The crucial understanding is that the coherence length over which the charge-conversion resistance occurs has a length shorter than the electron-electron scattering length. Therefore we need to analyse this charge-conversion resistance in terms of the distribution-function $f(E)$. However, beyond that length and given a length of the device larger than the electron-electron interaction length we can apply the concept of an electron temperature characterised by the temperature $T_e(x)$.    

\subsection{{\label{sec:NSN}}Normal equilibrium reservoirs}
In a recent experiment\cite{Shcherbatenko2016} we have studied a number of devices to separate experimentally the properties of the contacts from the properties of the NbN itself.  In practice, the contacts provide the boundary conditions for the driven superconducting state in the bare NbN film.  The width and length of the devices shown 
in Fig.~\ref{fig:3Dview} have been varied.  Table \ref{tab:Table1} shows the device parameters. We define, $T_{c1}$ as the superconducting transition temperature of the NbN of the active material itself, $T_{c2}$ of the NbN-Au bilayer, and $T_{c3}$ of  the NbN-Au-Cr-Au multilayer (Table \ref{tab:Table2}).

\begin{table}[t]
  \centering
  \caption{Parameters of the devices made of the same NbN film. Note the increase in the normal resistance per square with decreasing temperature with the maximum $R_p=R_{peak}$,  as well as the variations in $R$ and $T_{c1}$ from sample to sample. }
  \label{tab:Table1}
  \begin{tabular}{cccccc}
  \hline\hline
    Dev \# & W($\mu m$) & L($\mu m$) & $R_{300} (\Omega/\square)$ & $R_{p}(\Omega/\square)$ & $T_{c1}(K)$ \\
    \hline
  &  &  &   & &\\
  2 & 0.99 & 0.4 &688 & 1366&9.2\\
  7 & 2.01 & 0.4 & 650 &1276&9.5\\
 8 & 2.01 & 0.4 &685&1328&9.3\\
9 & 2.57 & 0.4  &628&1133&9.6\\
10 &  2.57 & 0.4  &621&1256&9.7\\
11 & 3.12 & 0.4 &601&1102&9.8\\
    \hline\hline
  \end{tabular}
\end{table}

\begin{figure}[t]
\begin{center}
\includegraphics[width=0.7\columnwidth]{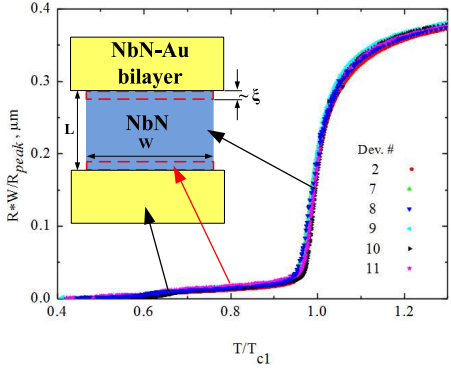}
\end{center}
\caption{\label{fig:RTDevices}Resistance as a function of temperature scaled on the normal state resistance just above $T_{c1}$ and multiplied by the width $W$ for the set of devices listed in Table \ref{tab:Table1}.  Inset: Schematic view of the center of the device. Black arrows indicate which part is determining the critical temperature, $T_{c1}$ for the NbN (blue) and $T_{c2}$ for the bilayer. The dashed red line indicates the superconducting resistive part of the NbN, which causes the plateau-resistance in the $R(T)$ trace indicated with the red arrow. {(Reproduced with permission from the authors \cite{Shcherbatenko2016})}}
\end{figure}

\begin{figure}[t]
\begin{center}
\includegraphics[width=0.68\columnwidth]{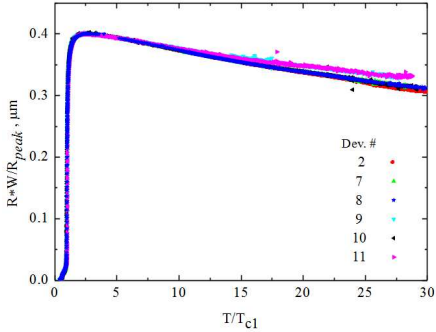}
\end{center}
\caption{\label{fig:SITcurves} Resistive transitions of the devices, all based on a single film of NbN, over a large temperature range. The data are rescaled to the same width and normalized to $R_{peak}$ (Table~\ref{tab:Table1}). {(Reproduced with permission from the authors \cite{Shcherbatenko2016})}} 
\end{figure}

Fig.~\ref{fig:RTDevices} shows a set of curves of the resistance as a function of temperature in a narrow temperature range. It shows the transition curve for uncovered NbN and the end transition of the NbN-Au bilayer (two black arrows). However, it is important to take into account a wider temperature range, which includes the resistance as a function of temperature to room temperature (Fig.~\ref{fig:SITcurves}). The latter curves clearly show a rise in resistance from room temperature to cryogenic temperatures.  The critical temperature $T_{c1}$, taken as the mid-point of the transition, marks the transition temperature of the uncovered NbN of the specific device. The values for $T_{c1}$ are given in Table \ref{tab:Table1}, and it is clear that they scatter. One also find a variation in the peak of the resistance just prior to the turn to superconductivity, listed also in Table \ref{tab:Table1}, abbreviated as $R_{p}$ for $R_{peak}$. This variation in $T_{c1}$ and $R_{peak}$ is a 
significant result because these devices are all made of the same film. 
The recent research on strongly disordered superconductors \cite{Sacepe2008,Pratap2013}  and summarised in Section \ref{sec:resistsuper} suggests that this variation is unavoidable. The reason is that the films, approximately 4 nm thick,  have a resistance per square of the order of  $1200 ~{\Omega/\square}$, which is equivalent to a resistivity of 480 $\mu\Omega$cm. The recent research has made clear that the competition between localization and superconductivity leads in these strongly disordered films, to a spatially fluctuating energy gap, even for atomically uniformly disordered materials \cite{Sacepe2008, Pratap2013}. Hence, superconducting properties need to be determined for each individual device to arrive at a consistent parametrisation in the comparison of the devices taking into account the variations in $T_{c1}$ and $R_{peak}$ shown in Table \ref{tab:Table1}. In 
Fig.~\ref{fig:SITcurves} the normal state resistances for all studied devices are shown scaled with the width $W$ and normalized to $R_{peak}$, with temperatures normalized to $T_{c1}$. Evidently, all curves are on top of each other and in agreement with the lithographically defined length of $0.4~\mu$m, as is evident from the vertical axis. The value of $R_{peak}$ in Table \ref{tab:Table1} will in the following be taken as the resistivity in the normal state of the superconducting film of the particular device, which will also be a measure of the elastic mean free path and the diffusion constant.

The evolution of the resistive superconductivity in the device is apparent from the resistive transition over a much more narrow temperature range around $T_{c1}$ and shown for all devices in Fig.~\ref{fig:RTDevices}. The observed resistance is multiplied by the width $W$ and divided by $R_{peak}$ to take out the dependence on the width and the dependence on the resistance in the normal state. We observe clearly two transitions,  a third more gradual transition is in this measurement not clearly discernible due to the noise. The fact that the normalisation to the NbN properties leads to an identical set of curves is a clear indication that we observe systematic intrinsic behaviour for all devices. Since the scaling is based on the width and resistivity of the NbN we must assume that the full stepwise resistive transition of these devices is due to the properties of the NbN and, as sometimes incorrectly assumed, not just only the transition at $T_{c1}$.   
\begin{table}[b]
  \caption{Device parameters relevant for the performance.}
  \label{tab:Table2}
  \centering
  \begin{tabular}{cccccc}
    \hline\hline
    Dev \# &$T_{c1} (K)$&$T_{c2} (K)$&$T_{c3} (K)$&$R_{pl}(\Omega \mu m )$&$V_c(mV)$ \\
    \hline
&&&&&\\
  2 & 9.2 &5.7  &4.3 &10.9 &1\\
  7 & 9.5 &5.3  &4.7  &12.1 &1\\
 8 & 9.3 &5.6  &4.6&13.1&1.05\\
9 & 9.6 &6.6   &5.5&12.9&1.35\\
10 & 9.7 &6.5   &5.3 &11.6&1.45\\
11 & 9.8 &6.1  &4.8&12.8&1\\
   \hline\hline
  \end{tabular}
\end{table}

With the values of $T_{c1}$ and $T_{c2}$ easily understood, we address the more difficult question of the origin of the observed resistance between $T_{c1}$ and $T_{c2}$.  It has an identical value for all devices if properly scaled on the geometric dimensions of the NbN and the NbN properties. In addition the device is, for that temperature range, an NSN device with the yellow parts in the inset of Fig.~\ref{fig:RTDevices} in the normal state and the blue part superconducting. As summarised in Section \ref{sec:conversion} a superconductor is resistive if charge is being converted from normal charge to Cooper-pair charge. The present data prove that in NbN the same phenomenon appears between $T_{c1}$ and $T_{c2}$.  For the devices listed in Table \ref{tab:Table1} the width $W$ was varied while the length $L$ was kept constant. The width of the NbN-Au bilayer was kept constant, as well as all other parameters. The plateau-resistance, $R_{pl}$, occurs between $T_{c1}$ and $T_{c2}$, when the NbN-Au bilayer is normal. The order of magnitude of the resistance is about 10 $\Omega$. As shown in Fig.~\ref{fig:RTDevices} all curves collapse onto one curve when scaled to the width $W$ of the NbN, while all other dimensions the same.  If it is due to the same mechanism as discussed in Section \ref{sec:conversion} the resistance should occur over a length of the order of the coherence length in the NbN. On the vertical axis the data are scaled to the normal state resistance of the device at the peak value just above $T_{c1}$, called $R_{peak}$. Horizontally the temperature is normalized to the $T_{c1}$ of the main resistive transition attributed to NbN.  Fig.~\ref{fig:RTDevices} clearly shows that  all the curves follow an identical trace, proving that indeed all the properties of the resistive transition are controlled by  the properties of the bare NbN.  From the sheet resistance of  1200 $\Omega/\square$ we find that the effective length of the two resistive parts of the superconducting NbN is about 8 nm.  This points to a coherence length of the order of 3 nm to 4 nm, in very good agreement with other estimates of the coherence length for NbN. We conclude that for bath temperatures between $T_{c1}$ and $T_{c2}$ the device can be viewed as an NSN device, with the observed plateau-resistance due to the nonequilibrium charge conversion length inside the uncovered NbN. It should be emphasized that this resistance is realized in the NbN although it is in the superconducting state.  Obviously this resistive contribution is expected to have  a negative effect on the mixing performance of NSN devices because it is only weakly dependent on changes in electron temperature.  
\begin{figure}[t]
\begin{center}
\includegraphics[width=0.68\columnwidth]{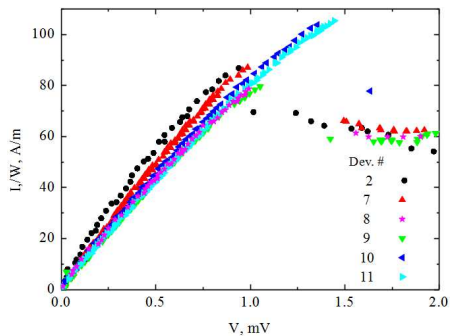}
\end{center}
\caption{\label{fig:ScaledIV}Current-voltage characteristics measured at $T/{T_{c1}}=0.8$ showing that they are nearly linear, indicative of the insensitivity of the conversion resistance to the power delivered to the system. All curves terminate at a specific point in the $I,V$ plane, which is interpreted as a critical voltage arising from the increased energy-mode nonequilibrium. {(Reproduced with permission from the authors \cite{Shcherbatenko2016})}}
\end{figure}

Having identified the origin of the resistance between $T_{c1}$ and $T_{c2}$ it is worth extending the analysis to a study of the current-voltage characteristics in the same temperature range. A typical set, measured at a normalized temperature $T/T_{c1}=0.8$,  is shown in Fig.~\ref{fig:ScaledIV}. The $I,V$ curves are almost linear and are terminated at some critical point in the $I,V$ plane, after which the system switches to the normal state. There is some variation from device to device with respect to this critical point, but if properly scaled they are quite similar and the linearity is not trivially expected. The linear behaviour indicates that the conversion-resistance at the entry and exit of the superconducting material does not change with increasing bias voltage. Apparently the extra energy which enters the system does not effect the charge-mode of non-equilibrium at the NS interface, which describes the conversion-resistance. However, in addition to the charge mode of nonequilibrium there is also an energy-mode of nonequilibrium, analogous to heating or cooling. This critical point  has been identified \cite{Keizer2006}  as a critical voltage at which the superconducting state becomes unstable for  a voltage approximately equal to $(1/2\sqrt{2})\Delta_0$, with $\Delta_0$ the equilibrium energy gap  of the superconductor. However, in this model \cite{Keizer2006, Vercruyssen2012} it is assumed that the length of the superconducting wire is short compared to the electron-electron interaction time $\tau_{ee}$, leading to the parameter range $\xi<L<\Lambda_{ee}$. Consequently, a position-dependent non-thermal 2-step distribution function occurs, like for normal metal wires studied in detail before \cite{Pothier1997}. For NbN this assumption is not justified because the electron-electron interaction time $\tau_{ee}$ is estimated to be 2.5 ps \cite{Annunziata2010}or 6.5 ps \cite{Ilin2000}. For the resistance per square of our samples the diffusion constant $D$ is estimated by rescaling the numbers of previous work from \cite{Semenov2001} to be 0.2 cm$^2$/s, leading to a characteristic length $\Lambda_{ee}$ of 7 nm to 12 nm. Hence, our NbN devices are in a regime where $\xi<\Lambda_{ee}<<L$, which justifies our interpretation of the charge-conversion process.  However, the energy mode of the distribution function has time to become thermal over the length of the superconductor. Hence, it is to be expected that for niobium-nitride an effective electron temperature \cite{Pothier1997} is given by $T_e(x)=\sqrt{T^2+x(1-x)V^2/L_0}$.  Here, $T$ is the temperature of the contacts, $V$ the applied voltage, and $L_0$ the Lorenz number.  The coordinate $x$ runs from 0 to 1 along the superconducting wire. It is to be expected that if $T_e$ is equal to $T_{c1}$ the device will become dissipative at the maximum temperature in the center at $x=1/2$. One expects therefore at $T/T_{c1}=0.8$ that $V_c=1.9~10^{-4} ~T_{c1}$ with $T_{c1}$ in K and $V_c$ in V in quite good agreement with the data. The fact that $V_c$ is slightly lower can be reconciled by taking into account the reduced heat-diffusion, due to the order parameter profile. In addition, some electron-phonon relaxation might also contribute. 

We may conclude that the DC properties of niobium-nitride hot-electron bolometers in the temperature range in which they constitute a NSN device can be understood as having an observable resistance due to charge-conversion processes and a critical point in the current-voltage characteristic where an effective electron temperature is reached equal to $T_{c1}$ of the niobium nitride.  

\subsection{{\label{sec:S'SS'}}Superconducting equilibrium reservoirs}

A typical current-voltage characteristic for a practically used hot-electron bolometers at a temperature where the electrodes have become superconducting is shown in 
Fig.~\ref{fig:OperatedHEB}. It resembles initially the behaviour discussed in the previous section (\ref{sec:NSN}) in the sense that it shows a finite slope terminated by a critical point at about a few hundreds of microvolts. The full set of current-voltage characteristics is measured at an operating temperature of 4.2 K for different levels of LO-power \cite{HajeniusASC2005}. Also indicated (blue star-symbol) the optimal level of pumping  and dc bias with, in this case, a best noise temperature of 1200 K. For the unpumped curve a rising current is observed with a finite slope until a critical point is reached. It is followed by an erratic curve, which is a time-averaged behaviour of relaxation oscillations in the circuit due to the negative resistance slope beyond the critical point \cite{Vernon1968,Skocpol1974}. Beyond this erratic range the I,V curve is stable again and can be understood as due to a normal hot spot. 

\begin{figure}[t]
\begin{center}
\includegraphics[width=1.1\columnwidth]{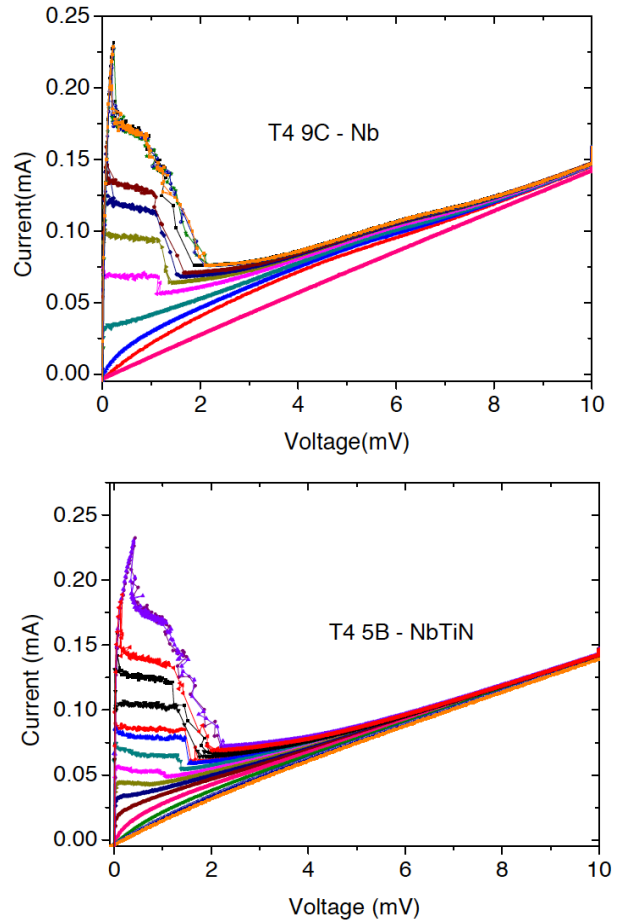}
\end{center}
\caption{\label{fig:Tarun}Current-voltage characteristics measured at an operating temperature of $4.2~K$ for different levels of LO-power \cite{Tarun2008}. The curves with the highest currents are without LO-power. Evidently these rising curves are for Nb contacts ($T_c=6$ K) different from the one for NbTiN ($T_c=10$ K). The NbN film has about $T_c=10$ K as well. {(Reproduced with permission from the authors \cite{Tarun2008})}}
\end{figure}

Although, a solid experimental basis has not yet been established we believe that the initial slope for zero LO-power is controlled by the same physics as for the temperature regime where the electrodes are normal. As a starting point in the analysis a difficulty is that the system should start with zero resistance because it is not a 2-point measurement of the superconductor like in the NSN case, but a S'SS' system. Therefore we can not start the analysis assuming a voltage-difference applied to a mesoscopic device. At present, a detailed model-study of a S'SS' device is not available. In addition this regime has hitherto drawn little systematic experimental interest by users of hot-electron bolometers, except for its termination labeled as 'critical current'.  However, in our view the curved line which evolves into a straight line with a slope which does not go through the origin, shown in 
Fig.~\ref{fig:Tarun} should be interpreted as a charge-conversion resistance as well.  

We assume that we can apply a voltage-difference to a S'SS' structure. Both S' electrodes are equilibrium reservoirs in a mesoscopic structure in the spirit of Section \ref{sec:pothier}. In the previous Section \ref{sec:NSN} this approach has been applied to NSN, meaning that all electrons in N with $E<\Delta$ contribute to the charge conversion resistance. With N replaced by S' and assuming a proximity-effected superconducting state, which can still be treated with a BCS density of states with a lower energy gap $\Delta'$,  only quasiparticles with energy $\Delta'<E<\Delta$ will contribute to the charge conversion resistance in S. In comparison to the NSN case the excitations are quasiparticles which have a charge $q$, which depend on the energy, being zero at gap-edge and approaching $e$ for higher energies. In addition, also a supercurrent can flow, although this supercurrent should undergo in the presence of a voltage-difference also a phase-slip process. The supercurrent as such does not contribute to the charge conversion resistance. So for each voltage we have a fraction of the current, which enters as 'normal' current and a fraction which enters as a supercurrent. This causes the shifted asymptote compared to the NSN case. Further theoretical and experimental work is needed to understand whether this intrinsic series resistance has a negative impact on the mixing properties and how it could be minimised. We will return to this point in Section \ref{sec:bandwidth}.

Finally, we emphasize that a critical current in a regular superconducting nanowire is a property of a moving Cooper-pair condensate \cite{Anthore2003,Romijn1982v1}, which is very much different from the critical voltage identified here for NSN and, as we expect also for S'SS'. The critical pair-breaking current is an equilibrium property, unrelated to the absorption of power. The critical voltage is due to the power fed into the quasi-particle system in a voltage-biased superconductor. Nevertheless, in this particular configuration of an S'SS' devices we assume that, initially, zero resistance will occur at the S'S interface. When a voltage is present it most likely evolves through a phase-slip process analogous to one-dimensional phase slip processes. 

\section{Distributed models of superconducting hot electron bolometer mixers under operating conditions}

\subsection{Distributed temperature model: hot-spot model}
Presently, the practical realisation of hot-electron bolometers has evolved towards devices consisting of a short, narrow piece of, most often, niobium-nitride thin films connected to normal electrodes, usually made of gold. Since gold overlaps partially the underlying niobium-nitride the device is a NSN device with a possible intermediate layer between N and S, which we label S'. In addition the device is driven by a dc current as well as by a THz current, from the local oscillator, to reach its operating point. Consequently, the operation of the device is a mixture of the non-equilibrium processes contained in the lumped element model and the position-dependent properties of the superconducting state as well as the heat balance, which contains a diffusive part and a part due to electron-phonon relaxation. Therefore, after the initial introduction of the principle of operation based on a lumped element analysis, it became critically important to find a model or at least a conceptual framework which dealt with the position-dependence of the temperature, of the superconducting order parameter and with the resistive properties. This complexity is further amplified by the fact that the relevant superconducting properties depend on the distribution function not just by the temperature but more precisely on the occupation of states over the energies on both sides of the Fermi-energy (called \emph{energy-mode} and \emph{charge-mode} non-equilibrium). Therefore, hot-electron bolometers are in principle very simple devices but they constitute a very complex interplay of very many different superconducting phenomena which depend on the 3 parameters:  position, time and energy simultaneously. And some of the relevant phenomena were, until recently, poorly understood. In addition the parameters of the superconductor itself are such that it is part of ongoing research on strongly disordered superconductors.           

\begin{figure}[t]
\begin{center}
\includegraphics[width=\columnwidth]{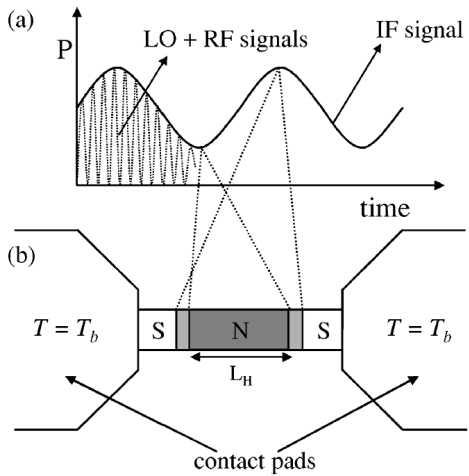}
\end{center}
\caption{\label{fig:Hotspot} Hotspot model in which the resistance is determined by the length of a normal domain $L_H$ embedded in a superconducting environment. The modulation of the electron temperature at the IF frequency is supposed to lead to a modulation of the length of the normal hot spot. The resistivity itself is the normal state resistivity which is temperature-independent. In addition it is assumed that at the interface between the superconductor and the normal metal there is also a charge imbalance type of resistance. {(Reproduced with permission from the authors \cite{WilmsFloet1999})}}
\end{figure}

Early attempts to model heterodyne mixing in superconducting HEBs have taken the hot spot model for a driven long superconducting wire \cite{Skocpol1974} as a starting point. It is designed to describe the current-voltage characteristics of short and long superconducting microbridges. Focusing only on the model for long microbridges the phenomenology can be summarized as follows.  For increasing current the zero-voltage state is terminated by the critical pair-breaking current, where the kinetic energy of the superconducting condensate exceeds the condensation energy. Beyond this critical pair breaking current a voltage-carrying state is found, which is close to the normal state resistance. For decreasing current a plateau in current develops, where the voltage rapidly decreases with decreasing current, and which marks the return switching current. To map the whole trajectory it is convenient to use a bias which acts as a voltage-source. It has been shown \cite{Skocpol1974} that the full voltage carrying state can be understood as due to a self-heating-maintained normal domain in an otherwise superconducting wire. In a companion paper the authors identify, close to $T_c$, a regime where this hot spot is preceded by a voltage-carrying state due to the emergence of phase-slip centers \cite{SkocpolJLTP1974} with a length given by twice the charge imbalance length $\Lambda_{Q^*}$ introduced in Section \ref{sec:hotsuper}. The evolution of these phase slip centers into a normal hot spot has also been modeled \cite{Stuivinga1983}.        

{Many attempts have been made to expand the hot-spot model to a level that it would capture the behavior of actual HEBs. Unfortunately, the hot-spot model has one fundamental assumption, which ignores the resistive properties of a superconductor as described in Section \ref{sec:resistsuper}. Rather than starting with the resistive properties of a superconductor it starts with the assumption that the resistance of the device is due to the resistivity of a normal metal with a temperature-independent} resistivity $\rho$ over a certain length $2x_n$ of the device (Fig. \ref{fig:Hotspot}).  This normal domain is sandwiched between superconducting domains, which have zero resistance. The effect of a modulation in THz power is to increase $2x_n$, the length of the normal domain, and therefore to increase the resistance and hence the voltage across the device . This approach has been introduced by Ekstr\"om \cite{Ekstrom1995} and further elaborated by {Merkel et al \cite{Merkel1999, Merkel2004}} and Wilms Floet et al \cite{WilmsFloet1999}. Although the work has been helpful in guiding the development of hot-electron bolometer mixers it {lost sight of the fact that at} the heart of the operation of hot-electron bolometers is the \emph{temperature-dependent} resistivity of a \emph{superconductor} (Section \ref {sec:resistsuper}) and \emph{not} the \emph{temperature-independent} resistivity of the normal state.

\subsection{{\label{sec:barends}}Distributed superconducting resistivity model}
An analysis of the optimal operating point for hot-electron bolometer mixers shows that the dc voltage is substantially below the voltage that one would expect for the NbN film in the normal state. In the previous section we have argued that it is reasonable to work with an electron temperature and a phonon temperature. This electron temperature determines the local resistivity, which should be temperature dependent in order to produce a measurable and sensitive IF signal. As discussed above the hot-spot model does not contain a great deal of sensitivity to variations in the electron temperature. Instead, one should assume  that the local resistivity is due to the emergence of free vortices as given by the Berezinskii-Kosterlitz-Thouless theory (Section \ref{sec:BKT}). This approach \cite{Barends2005} assumes that the local electron temperature generates locally the unbinding of the vortex-anti-vortex pairs of a superconducting film with a large penetration depth. The increased density of free vortices are the source of the resistivity in the superconductor, with the density of free vortices being position-dependent. The net magnetic field from the vortices is zero because it originates in the unbinding of vortex-anti-vortex pairs. So there should be an equal density of up-vortices compared to down-vortices. And the highest density should be at the highest temperature. From this perspective the local resistivity given by:

\begin{equation}
\label{vortex_pho}
\rho=\rho_n 2\pi \xi^2 N_f(J, T_e)
\end{equation}
which essentially takes the Bardeen-Stephen result \cite{BardeenStephen1965} for the resistance per vortex times the density of free vortices. We include in this expression, apart from the dependence on electron temperature, a dependence on the current. The current is providing the Lorentz-force on vortex-anti-vortex pairs leading to extra dissociation, and hence extra free vortices. It has been shown \cite{Benfatto2009}  that the BKT transition depends on finite-size effects and on inhomogeneities. It is therefore impossible to write down an equation with sufficient general validity. The general concepts have been very well documented and have been applied  to very thin NbN films \cite{KamlapureAPL2010} to analyse measurements of the penetration depth, which is a measure for the density of free vortices. 

\begin{figure}[t]
\begin{center}
\includegraphics[width=0.9\columnwidth]{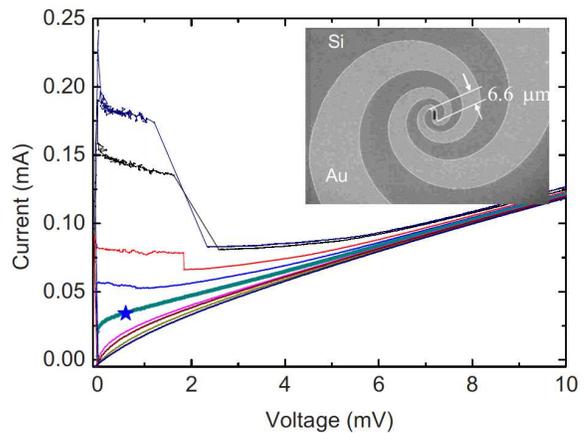}
\end{center}
\caption{\label{fig:OperatedHEB}Current-voltage characteristics measured at an operating temperature of 4.2 K for different levels of LO-power \cite{HajeniusASC2005}. Also indicated the optimal level of pumping and dc bias with, in this case, a best noise temperature of 1200 K. For the unpumped curve a rising current is observed with a finite slope until a critical point is reached. It is followed by an erratic curve, which is a time-averaged behavior of relaxation oscillations in the circuit due to the negative resistance slope beyond the critical point \cite{Vernon1968,Skocpol1974}. Beyond this erratic range the I,V curve is stable again and can be understood as due to a normal hot spot, where mixing is not effective. {(Reproduced with permission from the authors \cite{HajeniusASC2005})}}
\end{figure}

These results provide evidence that it is justified to consider the emergence of resistivity in thin films of NbN as used for hot-electron bolometers within the framework of the BKT-transition. Since, a full theoretical description is complicated by material properties, pinning sites, inhomogeneities, granularity, and finite-size effects, one uses instead an independently determined empirical relation for the resistive transition of the NbN film in the presence of a current. The intrinsic transition is measured separately for a part of the NbN film for different currents. At this point it is justified to use the empirical relation shown in Eq.~\ref{empirical rho(T)} to describe the empirically found resistive transitions. However, since the resistive transition is dependent on the transport current one uses for increasing currents a curve, which shifts to lower temperatures and the apparent downshift of the critical temperature Tc obeys the empirical relation
\begin{equation}
\label{empiricalRT}
\frac{I}{I_c}=\left(1-\frac{T_c(J)}{T_c(0)}\right)^\gamma
\end{equation}
which for the 4 nm thick films of niobium-nitride has been found to have a value for $\gamma$ of $\gamma=0.54$. 

The current-voltage characteristics for a hot-electron bolometer under operating conditions follows now from:       
\begin{equation}
\label{vortexIV}
V=J\int\rho(x, J, T_e(p_{rf},p_{dc}))dx
\end{equation}
with $\rho$ the empirically determined resistivity, which we assume to be also locally valid based on the local electron temperature and the overall current density. The temperature $T_e(x)$ is determined from a numerical calculation of the equations for the heat-balance for the electrons and for the phonons. A local electron temperature is determined for electrons (e) and phonons (ph), treating the system as if it were in the normal state,  
since the device is operated close to $T_{c1}$: 
\begin{eqnarray}
\frac{d}{dx}\left(\lambda_e\frac{d}{dx}T_e \right)+p_{dc}+p_{LO}-p_{e-ph}&=&0 \nonumber\\
\frac{d}{dx}\left(\lambda_{ph}\frac{d}{dx}T_{ph}\right)+p_{e-ph}-p_{ps}&=&0
\label{Heterodyne}
\end{eqnarray} 
Here, $\lambda$ denotes a thermal conductivity for electrons or phonons, $p_{dc}=J^2\rho(x)$ is the locally generated dc power. Since NbN is in the regime given in Table \ref{tab:Table3} the use of a local electron temperature $T_e(x)$ is a justified starting point. 
\begin{figure}[t]
\begin{center}
\includegraphics[width=\columnwidth]{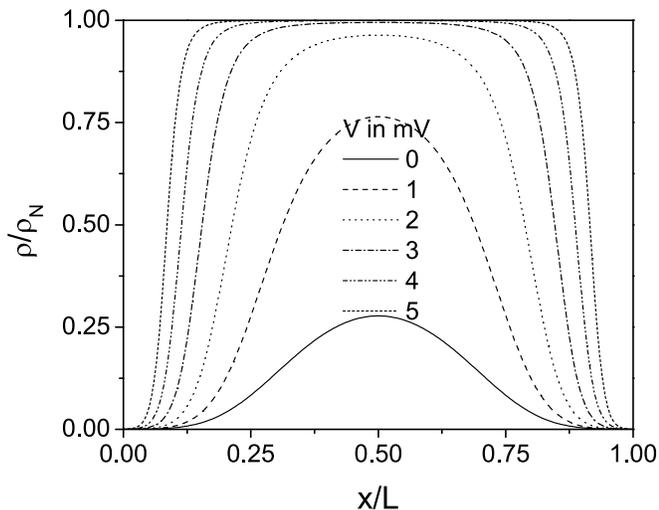}
\end{center}
\caption{\label{fig:rho_x} The resistivity as a function of position based on the electron temperature using the local resistivity as determined empirically from a separate measurements meant to have a set of data, which reflect the content of Eq.~\ref{vortex_pho}. {(Reproduced with permission from the authors \cite{Barends2005})}}
\end{figure}
In previous work \cite{Barends2005} the electron temperature is used to determine the local resistivity in the superconductor leading to a description of the current-voltage characteristic under operating conditions from:
\begin{equation}
V=J\int^{L/2}_{-L/2}\rho[x,J,T_e(x,p_{LO},p_{dc})]dx
\label{IV1}
\end{equation}
with $\rho$ the local value of the resistivity of the superconductor. For a range of voltages the result is shown in Fig.~\ref{fig:rho_x}. 

The empirically found operating point is between 0 and 1 mV, which corresponds to the full curve and the dashed curve. The full curve is the result for only rf power, whereas the shift to the dashed curve reflects the effect of the increase due to the dc power. The figure shows clearly that for higher dc powers the bridge gradually evolves indeed to a normal hot spot, where the resistivity is equal to the temperature independent normal state resistivity $\rho_n$. The region in which the HEB responds very effectively is in the regime where the resistivity is still very sensitive to the electron temperature.    

\begin{figure}[t]
\begin{center}
\includegraphics[width=\columnwidth]{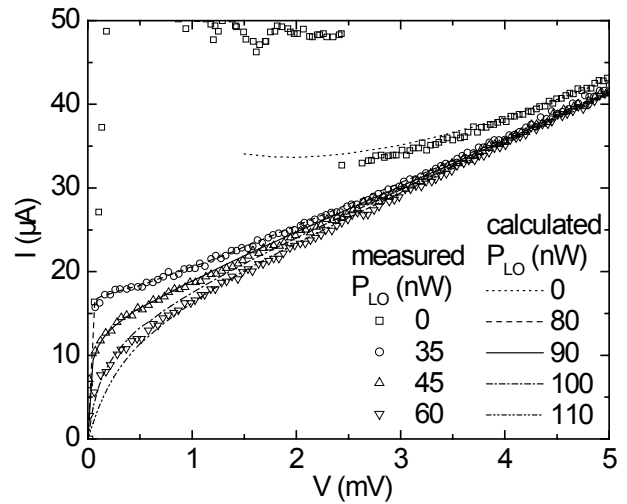}
\end{center}
\caption{\label{fig:BarendsIV} Experimental current-voltage characteristic for different rf power from the local oscillator (LO), compared to predictions from the local resistivity model, as follows from Eq.~\ref{vortexIV}.{(Reproduced with permission from Barends et al\cite{Barends2005})} }
\end{figure}

This analysis has been used to simulate the current-voltage characteristics under operating conditions. In Fig.~\ref{fig:BarendsIV} a set of measured current-voltage characteristics for different LO-powers is compared with the results of the simulation with excellent agreement. After determining experimentally for a NbN film Eq.~\ref{empiricalRT} for $T_c(I)$ and using it together with Eq.~\ref{empirical rho(T)} the local resistivity for a given electron temperature $T_e$  and a given current $I$ is known. This is used in Eq.~\ref{Heterodyne} to determine $p_{dc}$ together with Eq.~\ref{IV1} to determine the voltage $V$ for a given current $I$. The uniformly absorbed LO-power is used as an adjustable parameter, which shows a difference of about a factor of 2, which can only be accounted for analyzing more deeply the relation between applied and absorbed power at THz frequencies (in this case 1.6 THz was used). The good agreement between the experimental curves and the simulated current-voltage characteristics demonstrate that the interpretation of the resistivity of the hot-electron bolometers as being due to the \emph{resistivity of the superconductor} and the quantification in terms of an empirically determined relationship is very successful. And given the parameters applicable to NbN it proves that the operating point for optimal mixing results is due to a superconducting resistivity arising from thermally generated free vortices in analogy to the Berezinskii-Kosterlitz-Thouless phase transition in 2-dimensional films.      

\subsection{{\label{sec:bandwidth}}Distributed superconducting order parameter model: IF bandwidth}

\begin{table*}[t]
  \centering
  \caption{Parameter range for different materials assuming a device-length, $L$, of 400 $nm$,  to show that for different materials the physics is in different regimes (indicated by Range). The actual experimental values may vary depending on the deposition conditions, {temperature and resistance per square, like for example shown in detail by Santhanam and Prober for aluminium \cite{Santhanam1984}}.  The electron-electron scattering length is defined as $\Lambda_{ee}=\sqrt{D\tau_{ee}}$, and the coherence length $\xi$ given by $\sqrt{\xi_0 \l}$}
  \label{tab:Table3}
  \centering
  \begin{tabular}{cccccc}
   \hline\hline
    Material &$D (cm^2/s)$&$\xi(nm)$&$\tau_{ee}(ps)$&$\Lambda_{ee}(nm)$&Range \\
    \hline
&&&&&\\
  Al-1\cite{Boogaard2004,Vercruyssen2012}& 160   &124 &$\simeq 6~10^3$ &$10^4$&$\xi< L<\Lambda_{ee}$\\
  Al-2\cite{Siddiqi2002}& 10  &50 &$\simeq 6~10^3$ &$2.5~10^3$&$L\simeq \xi<\Lambda_{ee}$\\
 Nb\cite{Burke1996,Burke1999,Gershenzon1990}&  1&40&100&100&$L\approx \xi,\Lambda_{ee}$\\
NbN\cite{Shcherbatenko2016} & 0.5 & 4 &2.5-6.5&11-18 &$\xi<\Lambda_{ee}<<L$\\
   \hline\hline
  \end{tabular}
\end{table*}

In Section \ref{sec:barends} we have argued that it is justified, under operating conditions, to use a local electron temperature $T_e(x)$ over the length of the niobium-nitride hot-electron bolometer. The heterodyne detection can therefore be understood as a modulation of the electron temperature $T_e$ at the IF frequency. The next step is to connect the modulation in $T_e$ to a modulation in resistivity leading through Eq.~\ref{IV1} to a modulation in voltage $V$ at the IF frequency. In Section \ref{sec:barends} we identified the source of the resistivity in the superconducting state as being due to the position-dependent density $n_f(x)$ and movement  of free vortices due to the Lorentz force. In this analysis we have simplified the description by working with a local electron temperature and a local density of the vortices. In contrast in Section \ref{sec:hotsuper} we have emphasized that superconductors depend very sensitively on the distribution-function $f(E)$ and much more subtle than can be captured in a local electron temperature $T_e(x)$. This difference became clear in the study of the DC properties of NSN and S'SS' devices in the absence of LO-radiation, where the current-voltage characteristics are controlled by a charge-conversion resistance. This phenomenon can only be understood by using $f(E)$ and taking into account the energy-dependence of the superconducting properties. In 
Fig.~\ref{fig:NathanProfiles} solutions for the order parameter $\Delta$ are shown for the model-study based on aluminium-wires \cite{Vercruyssen2012}. Two different regimes are shown, which represent the local superconducting properties, which emerge from a solution of the Keldysh-Green's function approach including both the energy-mode and the charge-mode of the non-equilibrium state, being present simultaneously. To describe hot-electron bolometers under operating conditions one would have to carry out this program and add to it the time-dependence of the moving vortices, which are the cause of the resistance under operating conditions. Such a program has not been carried out and therefore we will use the understanding of the source of the resistance in the superconductor together with the model-analysis \cite{Vercruyssen2012} as inspiration for a qualitative description of superconducting hot-electron bolometers under operating conditions including the energy-dependence of the superconducting properties.

One important question for hot-electron bolometer mixers is the IF bandwidth. The value of a model, even a qualitative one, is that it can provide guidance to optimise the devices. The proposal of Prober \cite{Prober1993} suggested to focus on 'contact-engineering', whereas the earliest work \cite{Gershenzon1990} suggested to focus on materials with a faster energy-relaxation rate. The two approaches have led to an early distinction between phonon-cooled and diffusion-cooled devices, {with the emphasis on the effect on the IF bandwidth. Such an effect has been observed in experiments on Al, Nb and NbC HEBs \cite{Karasik1996b, Burke1996, Burke1999}. In the practically used HEBs, made from NbN, and modeled with the diffusion equations shown in Eqs.~\ref{Heterodyne}, the step towards diffusion-cooled devices has not provided its anticipated effect on the IF bandwidth. It has been found experimentally\footnote{J. R. Gao and N. Vercruyssen, unpublished results Kavli Institute of Nanoscience, Delft University of Technology} that changing the length of the practical devices, from 0.1 to 0.4 $\mu$m,} has no influence on the IF bandwidth, which suggests that the electron distribution which controls the superconductive resistive properties is not limited by diffusion but, most likely, by phonon-cooling. {Nevertheless, in all cases diffusion is needed to create the temperature profile, whereas for the IF bandwidth one needs to understand the temporal response of the superconducting resistivity in the center of the bridge.} In fact, in comparing results on NbN, with results on NbTiN \cite{Puetz2011} and MgB$_2$ \cite{NovoselovAPL2017} it appears that the phonon-cooling dominates the IF bandwidth. On the other hand an increase in the bath temperature leads apart from a  deterioration of the noise temperature to an increase in IF bandwidth \cite{Tretyakov2011, Tretyakov2016}. Therefore the challenge is to understand why diffusion is an important ingredient in all distributed models, whereas at the same time the IF bandwidth appears not to benefit from this diffusion but appears instead rather controlled only by the electron-phonon relaxation rate. 

A superconducting hot-electron bolometer is in principle a mesoscopic device \cite{BeenakkerVanHouten1991,Pothier1997}.  In interpreting the results demonstrated for HEBs of different materials under these bias conditions, we first need to take into account the length $L$ of the device in comparison to the characteristic lengths for the microscopic processes.  A summary of relevant length scales for different materials is given in Table \ref{tab:Table3}, mainly provided to facilitate a translation from aluminum results to an application to niobium-nitride devices. 

Aluminium has a very long coherence length and a very long electron-electron interaction length, of the order of $1~\mu$m {(depending on temperature and resistance per square \cite{Santhanam1984})}. Hence, one can easily be in the regime $L<<\xi,\Lambda_{ee}$, like in the case of hot-electron bolometers studied by Prober \cite{Prober1993} and Siddiqi et al \cite{Siddiqi2002}. One can also perform experiments in the regime $\xi<L<\Lambda_{ee}$ \cite{Boogaard2004, Keizer2006, Vercruyssen2012}. For niobium one has usually an intermediate regime with $\xi$ of about 40 nm and $\Lambda_{ee}$ of 100 nm \cite{Gershenzon1990}, which was studied by Burke et al \cite{Burke1996,Burke1999}. For NbN, the coherence length over which the conversion resistance, discussed above, occurs is about 4 nm, shorter than the electron-electron scattering length, $\Lambda_{ee}$, of about 11 to 18 nm. The NbN device itself is 0.4  $\mu$m long and therefore the physics is controlled by the inequalities: 
\begin{equation}
\xi<\Lambda_{ee}<<L
\label{inequality}
\end{equation}
with $L$ the length of the bare NbN. Since the best mixing results to date have been achieved with NbN, we focus on the conditions valid for this material.

\begin{figure}[t]
\begin{center}
 \includegraphics[width=\columnwidth]{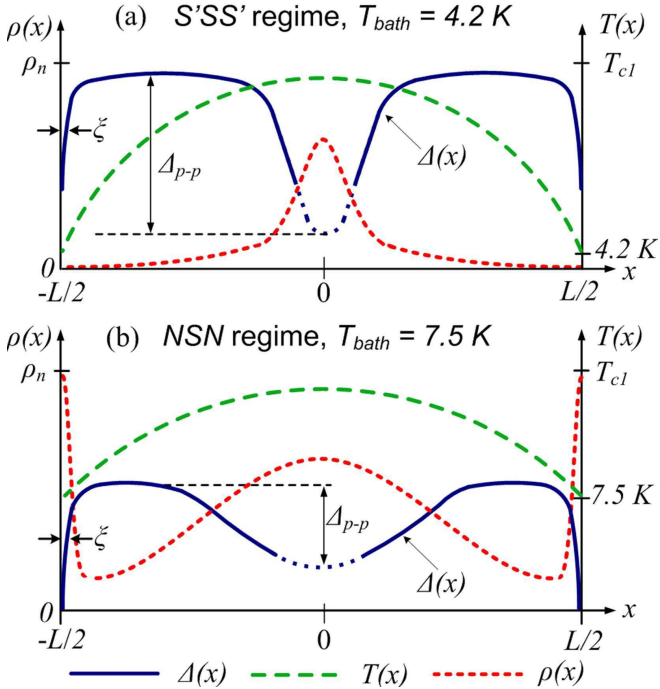}
\end{center}
\caption{\label{fig:profiles}Temperature, superconducting order parameter and local resistive superconductivity of NbN hot-electron bolometers under operating conditions at two different bath temperatures.  $T_{c1}$ and $\rho_n$ indicate values for the plain NbN (uncovered by a normal metal). (a) At bath temperature of $4.2\,K$ (S'SS' regime). In this figure the bias conditions are for an effective resistance of about 10\% of the normal state resistance. Optimal operating conditions are close to 50\%\cite{Hajenius2006} for which the extent of the resistive part along the bridge is larger\cite{Barends2005}. (b) At a bath temperature of $7.5\,K$ (NSN regime). {(Figures made with assistance of M. Shcherbatenko, I. Tretyakov and Yu. Lobanov.)}}
\end{figure}

In Fig.~\ref{fig:profiles} we have sketched two sets of curves for superconducting hot-electron bolometers under operating conditions. Panel b is for a bath temperature between $T_{c1}$, the critical temperature of the NbN film and $T_{c2}$ the critical temperature of the bilayer which forms the contact. In this regime the contact to the NbN is in the normal state and the device is a NSN device. Panel a is for a bath temperature below $T_{c2}$, which means that the contact is superconducting. The device can be labeled as S'SS', with S' a superconductor with a lower $T_c$ and a lower energy gap $\Delta$ than in S. In both regimes we assume that the electron temperature has to a first approximation a parabolic temperature profile, green dashed curves, like discussed in Section \ref{sec:barends}. In the center it approaches $T_{c1}$, at the edges they are at the bath temperature, because we assume that the contacts are electronic equilibrium reservoirs. 

Additionally, we have also sketched in both cases the distributed resistivity. In Section \ref{sec:resistsuper} the various ways resistivity can appear in a superconductor have been discussed. We distinguish the charge conversion resistance $\rho_c$ and the resistance due to time-dependent changes of the phase $\rho_\phi$. We have argued in Section \ref{sec:barends} that in the center the resistivity is due to movement of free vortices emerging from the Berezinskii-Kosterlitz-Thouless phase transition. Since passage of a vortex between 2 points is the analogue of a phase-slip event we call this process due to a time-dependent change of the macroscopic phase-difference, essentially like the Josephson-effect. In Panel b of Fig.~\ref{fig:profiles} the red dashed curve represents the two contribution to the resistivity. At the edges over a length of the coherence length the charge conversion resistance $\rho_c$ and the remainder represents $\rho_\phi$, which follows the profile of the local electron temperature, although in principle it should take into account the exponential rise in resistivity given by Eq.~\ref{eq:nelsonhalperin} or analogous to the empirical approach used in Section \ref{sec:barends}. In Panel a of Fig.~\ref{fig:profiles} only $\rho_\phi$ has been drawn, because we assume that the charge-conversion resistance $\rho_c$ does not play a role if the contacts are in the superconducting state. (This assumption is still subject to debate in view of the DC properties in the S'SS' case discussed above, but we expect it to play a minor role in the discussion about the IF roll-off.)      

The third curve (blue) in both Panels of Fig.~\ref{fig:profiles} represents the superconducting order parameter $\Delta(x)$. In a uniform superconductor it represents also the energy gap in the excitation-spectrum (the BCS-gap). For a system with spatial dependencies states below the energy-gap can occur and therefore we call it the order-parameter, indicating that superconducting correlations and a Cooper-pair condensate are present, but the density of states will also vary with position (some details of these aspects have been studied in aluminium \cite{Boogaard2004, Keizer2006, Vercruyssen2012}. The blue curves indicate that at the edges the order parameter is zero, for NSN, or finite but small for S'SS' and rises over the coherence length. Because of the temperature profile $\Delta$ will have the value of the bulk at that temperature and then decreasing upon approaching the center, because at this point the temperature is close to $T_{c1}$. Therefore at the center the order parameter has a minimum. Around the minimum the blue curve is dashed to indicate that here the value changes in time due to the passage of vortices. In principle, a good representation requires also the dimension perpendicular to the plane, but since this only effects the central part, the dashed curve is meant to represent the projection of the flux flow related changes in $\Delta(x)$ on a 2-dimensional plane.     

In Section \ref{sec:hotsuper},  Section\ref{sec:NSN} and Section\ref{sec:barends} it was emphasized that for the charge conversion resistance one has to describe the driven state in terms of the distribution function $f(E)$, because the coherence length $\xi$ is shorter than the electron-electron interaction length $\Lambda_{ee}$. Over the longer length scale it was adequate to take a local electron temperature $T_e(x)$. However, The profiles for the order parameter $\Delta(x)$ bring into the field of view that the qualitative description becomes inconsistent. In using Eq.~\ref{Heterodyne} the diffusion of hot electrons is not energy-dependent, whereas in Fig.~\ref{fig:profiles} there is obviously a valley in the order parameter, which blocks out-diffusion of electrons or quasiparticles for $E<\Delta_{p-p}$ (defined in Fig.~\ref{fig:profiles}). Therefore Eqs.~\ref{Heterodyne} should be replaced by an energy-dependent set of diffusion equations for superconductors. These are available and have been used for aluminium \cite{Vercruyssen2012}. At present such an analysis has not been carried out yet for niobium-nitride based hot-electron bolometer mixers.

The total power, $p_{dc}$ and $p_{LO}$, is absorbed by the device, which determines in principle $f(E,x)$, but in a way which depends on the length scales given in Table \ref{tab:Table3}. As follows from Table \ref{tab:Table1} for NbN for which the inequalities of Eq.~\ref{inequality} applies the two resistive contributions, $\rho_c$ and $\rho_\phi$, are spatially separated and the integrand in Eq.~\ref{IV1} can be split into two parts, leading to: 
\begin{equation}
V=J\int^{L/2}_{-L/2}[\rho_c(f(E,x))+\rho_\phi(x,J,T_e(x,p_{LO},p_{dc}))]dx
\label{IV2}
\end{equation}
The first term $\rho_c$ leads, in essence, to a resistance per unit width of $\rho_n\xi$, with $\rho_n$ the normal state resistivity. It is weakly dependent on the external parameters and is taken to be not dependent on $T_e$, as suggested by the prior work on Al \cite{Keizer2006, Vercruyssen2012}. The second contribution to the resistivity is in the usual way connected to the emergence of resistivity in a superconductor. Both contributions are illustrated in Fig.~\ref{fig:profiles} (red dashed curve). For other materials than NbN these two contributions are spatially intermixed and can not be split into two parts. We focus further on the  practically used NbN mixers.

The heterodyne mixing process on the level of Eqs.~\ref{Heterodyne} and Eq.~\ref{IV2} can be understood as follows. The LO-power and the signal power are uniformly absorbed over the length $L$, because $\hbar \omega_{LO}, \hbar \omega_s>2\Delta$. These two coherent signals create the heterodyne signal  of $\sqrt{p_{LO}p_{s}}\cos(\omega_{IF} t)$ at $\omega_{IF}$, which means that in Eqs.~\ref{Heterodyne} the term $p_{LO}$ should be supplemented with  $\sqrt{p_{LO}p_{s}}\cos(\omega_{IF} t)$. Consequently, the solution of the set of equations, Eqs.~\ref{Heterodyne} is $T_{e}(x,t)=T_e(x)+\delta T_e(x)\cos(\omega_{IF} t)$.  Hence, heterodyne mixing manifests itself in an oscillatory component of the \emph{electron temperature}, at a frequency $\omega_{IF}$, over the full length of the device, although the amplitude will vary over the length. The remaining question is how this oscillatory electron temperature is converted into an observable voltage signal through  
Eq.~\ref{IV2}. As stated above $\rho_c$ is very weakly dependent on $T_e$ and therefore has a negligible contribution to the IF-signal in Eq.~\ref{IV2}. To establish the connection between $T_e(x,t)$ and the resistivity $\rho_\phi$ we need to reconsider the description on the level of a Fermi-Dirac distribution with an electron temperature $T_e(x)$. For the 2nd term in the integrand of Eq.~\ref{IV2} a thermal model is assumed appropriate for NbN because $L>>\Lambda_{ee}$. However, as is evident from Fig.~\ref{fig:profiles} the order parameter has a minimum at $x=0$ and a \emph{non-thermal} energy distribution is likely to arise. Therefore Eq.~\ref{IV2} becomes

\begin{eqnarray}
V=J\int_0^\infty  \int^{L/2}_{-L/2} \bigl[\rho_c(f(E,x))+~~~~~~~~~~~~~~\\
~~~~~+\rho_\phi(x,J,f(E,x,p_{LO},p_{dc}))\bigr]dx dE
\label{IV3}
\end{eqnarray}
in which the dependence on $T_e(x)$ has been replaced by the dependence on $f(E,x)$. The message is that a Fermi-Dirac distribution characterised by an electron temperature is no longer adequate because of the rate of change of the order parameter compared to the electron-phonon relaxation time.   
\begin{table*}[t]
  \centering
  \caption{Relaxation times relevant for IF bandwidth for devices of $0.4~\mu m$ length. D in $cm^2/s$, $\tau_D=L^2/\pi^2D$ in $ns$, $\tau_{ee}$ the electron-electron interaction time  in $ps$ and $\tau_{e-ph}$ the electron-phonon interaction time in $ns$. The given relaxation times are typical values, in practice depending on film thickness and deposition conditions.}
  \label{tab:Table4}
  \centering
  \begin{tabular}{cccccc}
    \hline\hline
\newline
    Material &$D (cm^2/s)$&$\tau_D(ns)$&$\tau_{ee}(ps)$&$\tau_{e-ph}(ns)$ \\
    \hline
&&&&&\\
  Al-2 \cite{Siddiqi2002}&10& 0.016  &$\simeq 6~10^3$ &$20$ &\\
 Nb \cite{Burke1996,Burke1999,Gershenzon1990}& 1&0.16&100&1&\\
NbN \cite{Semenov2009,Gousev1994} &0.5&0.3 &2.5-6.5&0.01-0.02 &\\
    \hline\hline
  \end{tabular}
\end{table*}
The rate of change of the order parameter at the center of the superconducting microbridge is controlled by the flux-flow vortices, which are created due to the Berezinskii-Kosterlitz-Thouless phase transition. The process of flux flow was soon after the discovery of the Josephson-effect understood theoretically as being closely related \cite{Anderson1965,Kulik1966} and supported by experiments \cite{Fiory1971}. The movement of Abrikosov vortices can interact with external radiation producing the Shapiro steps familiar from the Josephson-junction. The passage of a vortex between two points in a superconductor means a change of the phase-difference, $d\phi/dt$, which because the universality of the Josephson relation, means a voltage $V=\hbar/2e~d\phi/dt$. This insight was applied  \cite{SkocpolJLTP1974} to narrow superconducting films, as well as superconducting microbridges \cite{likharev1979}. It implies that in hot-electron bolometers with an optimal operating point at about 700 $\mu$V \ref{fig:OperatedHEB}, the rate of change is approximately 350 GHz. Which means that at the center at each point perpendicular to the direction of the transport-current the order parameter changes with a rate given by this frequency.  This frequency is much faster than the modulation of the electron temperature at the intermediate frequency and can be converted into an appropriate time-averaged quantity. The remaining question is how this time-averaged quantity depends on the electron temperature or the distribution-function.

In a uniform model \cite{PerrinVanneste1983} the response of the superconductor to optical radiation has been studied by focusing on the electron and phonon temperature, as well as the density of excess quasiparticles. Instead we propose to focus not on the temperature but on the non-thermal distribution function $f(E)$ and in addition not on the quasiparticle density but rather on the observable the resistance of the superconducting device as expressed in Eq.~\ref{IV3}. As suggested by Fig.~\ref{fig:profiles} we therefore split the distribution-function at the center of the hot-electron bolometer into two parts. For energies $E>\Delta_{pp}$, called $f_h(E)$, the electrons can diffuse sideways and can relax by electron-phonon relaxation. For energies $E<\Delta_{pp}$ the population of electrons, $f_l(E)$,  can only relax by electron-phonon processes and not by diffusion. In both cases the relaxation to the phonon-bath may depend on an increased phonon-temperature and be limited by phonon-escape. This will lead to a quantitative difference but does not effect the conceptual difference between $f_h(E)$ and $f_l(E)$. Both populations will be driven by the power from the heterodyne detection-process and develop a modulation $\delta f_{h,l}$ at the IF frequency. It is to be expected that the relevant relaxation time for $\delta f$ is shorter or equal for $\delta f_h$ compared to $\delta f_l$. The latter is predominantly controlled by the electron-phonon relaxation processes. The observable, the resistive properties in the superconductor is expressed in Eq.~\ref{IV3}. We expect that $\rho_c$ can be made negligible, the crucial quantity is $\rho_\phi$, in which the dependence on $f(E)$ is indicated but without a distinction between $f_h$ and $f_l$. Since the properties of a superconductor are most sensitive to the low-lying quasiparticle states we expect that the dominant contribution from $f(E)$ to $\rho_\phi$ will come from $f_l(E)$. This statement then leads naturally to the conclusion that the IF bandwidth of hot-electron bolometer mixers is determined by $f_l(E)$ and limited electron-phonon relaxation processes, including the phonon-escape time. Therefore it is beneficial to search for other materials and it is understood why in practical devices a change in the length has no effect on the IF bandwidth.     

Experimentally the IF bandwidth indicates a relaxation-time of the order of a few GHz. The rate of change of the order parameter due to flux flow is estimated to be in the hundreds of GHz range. Table \ref{tab:Table4} shows a number of values relevant for a discussion of the IF bandwidth. Since, we focus on practical hot-electron bolometer mixers we focus on NbN. As shown in Eq.~\ref{vortex_pho} the resistivity depends on the density of free vortices as well as the contribution of each vortex to the resistivity. Both are dependent on the electron temperature $T_e$ or if, non-thermal, on the specific distribution-function $f(E)$.  To reach a complete understanding of heterodyne mixing in superconducting hot-electron bolometers we need to connect $T_{e}(x,t)=T_e(x)+\delta T_e(x)\cos(\omega_{IF} t)$ to the local resistivity, because this causes the voltage signal at $\omega_{IF}$. At present, it is unclear whether the full answer to this question is in the temperature (or non-equilibrium) dependence of the density of free vortices or in the temperature dependence of each individual vortex to the resistivity.          

\section{Concluding remarks}

The material collected in this review forms, to the best of our knowledge, the context within which a full understanding of the simple and useful superconducting hot-electron bolometer mixers devices must be developed. 
\begin{itemize}
\item{The DC current-voltage characteristic in the absence of LO-power should be more systematically studied and interpreted as partially due to a charge-conversion resistance which depends on the nature of the contacts, which supply the electrons or quasiparticles to the superconducting film of NbN.}
\item{In our view there is at present not enough systematic and insightful experimental research done on the contribution of the charge-conversion resistance  to the mixing properties if it is present, how it can be minimised.}
\item{A model-study, both theoretical and experimental,  of the charge conversion resistance between a superconducting wire connected to superconducting electrodes with a lower energy gap would be very useful.}   
\item{A systematic analysis of the critical point in the current-voltage characteristic without LO-power in relation to the superconducting properties and the energy-mode of non-equilibrium would further reveal the information that can be extracted from this quantity to characterise the hot-electron bolometer mixers.}
\item{It would be advisable to study the resistivity in a confined geometry of a thin superconducting films with a high normal state resistivity to determine its dependence on the electron temperature.}
\item{An interesting problem to address is the dependence of the flux flow resistance on a time-dependent non-equilibrium distribution-function}
\item{Based on the results, presented in this critical review, an increase in IF bandwidth is to be found in new materials with a fast electron-phonon relaxation rate{, as well as in a fast phonon-escape time.}}
\item{Given the origin of the resistive superconducting state used for heterodyne mixing the vortex-dynamics should be studied as a source of noise.}
\end{itemize}

\section*{Acknowledgments}
Many discussions and collaborations with a great number of colleagues and friends are gratefully acknowledged, but we would like
to thank especially, {R. Barends, G.R. Boogaard, J. R. Gao,  G. Goltsman,  R.S. Keizer, Yu. Lobanov, P. P\"utz, M. Shcherbatenko, I. Tretyakov, N. Vercruyssen, and M.-P. Westig}. 
This work was carried out with the support of the Ministry of Education and Science of the Russian Federation, contract 14.B25.31.0007 of 26 June 2013 and Grant no. 17-72-30036 of the Russian Science Foundation. TMK also acknowledges the financial support from the European Research Council Advanced grant no. 339306 (METIQUM).
\bibliography{HEBReview}

\end{document}